\documentclass[
prd,
preprint,
superscriptaddress,
tightenlines,
showpacs,
showkeys,
nofootinbib,
aps,
amsfonts,
amssymb,
]{revtex4}

\usepackage{color}
\usepackage{multirow} 
\usepackage{amsmath}
\usepackage{graphicx}
\usepackage{epic}
\usepackage{eepic}
\usepackage{epsfig}
\usepackage{latexsym}
\usepackage{float}
\usepackage[all]{xy}
\usepackage{amsfonts}
\usepackage{amssymb} 
\usepackage{color}
\usepackage{multirow}
\usepackage{tikz} 
\usepackage[colorlinks, urlcolor=black, citecolor=red, link
color=blue]{hyperref}

\usepackage[caption=false]{subfig}

\begin{document}

\title{Graviton Signals in Central Production at the LHC}

\author{Rahul Basu\footnote{Deceased}}
\affiliation{The Institute of Mathematical Sciences, Chennai, TN 600113, India}
\author{Tanumoy Mandal}
\email{tanumoy@imsc.res.in}
\affiliation{The Institute of Mathematical Sciences, Chennai, TN 600113, India}

\begin{abstract} 
In this paper, we study central production, in the inclusive case, to look
for graviton signals in large extra dimensional model using dilepton and diphoton channels. We carefully
analyze signal and possible Standard Model background 
processes and study the feasibility of such new physics searches in 
a relatively clean environment as in central production where the proton
fragments are mostly emitted in the forward direction, and there is a 
clear rapidity gap between them and the centrally produced system.
Our analysis shows that the LHC with 14 TeV center of mass energy and 100 fb$^{-1}$
(300 fb$^{-1}$) of integrated luminosity can probe the effective gravity scale
up to 3.6 TeV (4.4 TeV) in both the dilepton and diphoton channels.
 
\end{abstract}


\maketitle 

\section{Introduction}
\label{section:introduction}

The Standard Model (SM) has been extremely compatible with all the experimental
tests so far. However, there  are many theoretical attempts to go beyond the SM 
in order to answer some fundamental issues in the SM namely the hierarchy problems,
number of generations etc. Theories with extra spatial dimensions 
have been proposed to answer the hierarchy problems of the SM \cite{ArkaniHamed:1998rs,Randall:1999ee}. In what follows, in this paper we
look for the graviton signals as predicted in the large extra dimensional model (proposed by Arkani-Hamed, Dimopoulos and
Dvali \cite{ArkaniHamed:1998rs}) in the central diffractive processes at the LHC.

Forward Physics has been used with great success in studying the Standard Model (SM) 
physics at colliders. In particular, HERA (Hadron-Electron Ring Accelerator) has used forward detectors to greatly
enhance our knowledge of perturbative Quantum Chromodynamics (QCD) and its effects
in deep 
inelastic scattering (DIS) experiments. Moreover, the observation of diffractive events, 
those with large rapidity gaps between the final state particles, have been one 
of the most interesting aspects of HERA physics.

In the last decade the importance of forward physics has been recognized for 
LHC (Large Hadron Collider) physics too. This has resulted in a proposal to install forward particle 
detectors at ATLAS
and CMS in order to detect protons emitted in the forward direction, implying the 
existence of a centrally produced system with a rapidity gap between it and the 
protons~\cite{Royon:2007ah,Albrow:2008pn}. 
More
specifically, we are talking of a process of the kind $p + p \to p + M + p$, also 
known as Central Exclusive Production (CEP). In this case, the outgoing protons 
are emitted with
barely any loss of energy (typically less than 2\% of their longitudinal momenta) 
along with a centrally produced system $M$ with a rapidity gap between it and the 
outgoing
protons on either side. This is a clean production of the central system since 
it corresponds to just two forward protons in the final state along with the 
decay products used
to identify the central particle $M$ and no hadronic activity between them. This 
has therefore been described as being like a
vacuum fluctuation as the two 
protons pass each
other, producing the central particle.
Some good review on the 
phenomenology of exclusive processes can be found in \cite{HarlandLang:2012qz,Albrow:2010yb,Martin:2009ku}. 
Recently, in \cite{Ryutin:2012np} a general framework of exclusive double 
diffractive processes and its prospects are discussed.

One of the main aim of the forward physics at the LHC is to look for the exclusive Higgs production events. First in \cite{Dokshitzer:1991he}, it was pointed out that Higgs can be produced
in diffractive processes with rapidity gaps on either side at hadron-hadron 
colliders. Later this interesting scenario has been discussed in a series of 
papers~\cite{Fletcher:1993ij,Lu:1994ys,Lungov:1995iq,Cudell:1995ki,
Khoze:1997dr,Khoze:2000cy,Cox:2001uq,Petrov:2003yt,Forshaw:2005qp} by several authors. Recently, 
to compute the background for Higgs searches, in \cite{Maciula:2010vc} authors
have estimated the exclusive $b\bar{b}$ pair production cross section at the LHC.
In exclusive processes Higgs would be 
produced almost at rest through gluon-gluon fusion. 
The final 
state contains two protons and a Higgs and, since the process is exclusive, the
invariant mass of the Higgs is 
directly related to the energy loss of the outgoing protons \cite{Albrow:2000na}.
In exclusive production, measuring the energy loss of the protons one can determine the mass of the central system without looking at its decay products.

It was discussed in \cite{Khoze:2000jm}, if the outgoing protons scatter through small angles, the 
two-gluon system is in a $J_z=0,$ C-even, P-even state, where the $z$-axis is the
proton beam axis. This means that any new resonance must carry the $0^{++}$ 
quantum number in CEP.  In other words, this enables a clean determination of the 
quantum numbers of
any new resonance \cite{Kaidalov:2003fw}.

With the possibility of tagging the forward outgoing protons, the LHC can be 
turned into an effective gluon-gluon, photon-proton and photon-photon collider, 
giving rise to 
a major QCD and electroweak physics program~\cite{Khoze:2002nf,Cox:2004rv}.  
In addition, the possibility of producing quarkonium states like $\chi$, $J/\psi$, is 
something being envisaged at the ALICE detector. As discussed in \cite{Lebiedowicz:2011nb,HarlandLang:2011qd}, the measurement of the two-body decay channels of $\chi_{c0}$ to light mesons in exclusive process can be used to study the dynamics of heavy quarkonia and to test the QCD framework of CEP.

However all these studies require a QCD based model that allows us to couple the 
di-gluon system to the protons. The preferred model to describe this is the so-
called Durham
Model which we briefly describe in the next section. Before we do that however, 
we should point out that various other processes in the CEP scenario have 
recently been observed 
in the Tevatron at CDF which have given physicists the confidence that the Durham 
model is on the right track. Even though the Durham model was proposed almost a 
decade ago, it 
was not until 2006 that CDF saw diphoton production in the central rapidity 
region which was consistent with the gluon-gluon fusion Durham model~\cite{Aaltonen:2007am}. 
Subsequently CDF also found 
another exclusive process - $e^+e^-$ and $\mu^+\mu^-$ production -- produced by 
two photons~\cite{Abulencia:2006nb}. Here the Tevatron (and in future the LHC) acts as a photon photon 
collider. These 
processes also allow the calibration of the forward proton detectors proposed by 
CMS and ATLAS. Quarkonium production of $J/\psi$ and $\psi$(2S) as well as $\chi_c^0$ 
have also been 
observed through their decay products~\cite{Aaltonen:2009kg}. Exclusive dijet production has 
been observed, at CDF~\cite{Aaltonen:2007hs} and D0 \cite{Abazov:2010bk} in line with the predictions of the Durham model. 
These processes 
already seen at the Tevatron have given confidence that we can use digluon 
fusion is accordance with the Durham model to observe the signatures of new 
physics.

It was indicated in \cite{Khoze:2001xm} that new physics like SUSY,
extra dimensions etc. can be searched for in exclusive processes.   
In the exclusive configuration various BSM signatures namely
MSSM Higgs \cite{Heinemeyer:2007tu}, charged Higgs \cite{Enberg:2011qh}, triplet
Higgs \cite{Chaichian:2009ts}, radions \cite{Goncalves:2010dw,Kisselev:2005xn},
massive gravitons \cite{Kisselev:2005xn}, long lived gluinos \cite{Bussey:2006vx},
effects of quantum gravity \cite{Kisselev:2005tp} etc. have been discussed in the
literature. In this work we study inclusive double diffractive process as a
probe of large extra dimensions (ADD model) using dilepton and diphoton channels
via the exchange of KK graviton. 

In Sec. \ref{sec:ADDintro} we briefly introduce the ADD model.
In Sec. \ref{sec:diffrac_process} we discuss diffractive processes and summarize the Durham model. 
In Sec. \ref{sec:diffrac_produc} we consider dilepton and diphoton production via KK
graviton. We discuss about important SM backgrounds in Sec. \ref{sec:SM_bkgrnd}. In Sec. \ref{sec:numerical_result}  we summarize our numerical results and present LHC discovery potential of extra dimensions. Finally, in
Sec. \ref{sec:conclusion} we offer our conclusions.

\section{ADD Model}
\label{sec:ADDintro}

In this section we give a very brief introduction of the ADD model
\cite{ArkaniHamed:1998rs}.
This model assumes that space is $4+n$ dimensional where $n$ is the number of
extra spatial dimensions. All the SM particles are confined to the usual $(3+1)$D spacetime which is called ``brane''. Whereas, gravity can propagate in the 
extra dimensions. In this model, the $4$D Planck scale ($M_{Pl}\sim 10^{19}$ GeV) is a derived scale which is related to the fundamental Planck scale ($M_{S}\sim$ TeV) by
\begin{equation}
M_{Pl}^2= V_n {M_{S}}^{2+n};~~V_n = (2\pi R)^n
\end{equation}
where $V_n$ is the extra dimensional volume and $R$ is the compactification radius.
An important consequence of the ADD model is the appearance of a tower of
Kaluza-Klein (KK) modes as a solution to the linearized Einstein equation in 
$4+n$ dimensions. These KK modes are almost degenerate and separated in mass by
$\mathcal{O}(1/R)$ ($1/R$ $\sim 10^{-4}$ eV to $\sim 100$ MeV for $n=2-7$) terms. After KK decomposition, we have massive spin-2 
KK gravitons ($h^{k}_{\mu\nu}$) which have interactions with the brane localized SM states via
the energy momentum tensor $T^{\mu\nu}$ of the SM
\begin{equation}
\mathcal{L} = -\frac{\kappa}{2}\sum_k T^{\mu\nu}(x)h^{k}_{\mu\nu}(x)
\end{equation}
where $\kappa = \sqrt{16\pi}/M_{Pl}$ and the summation runs over all KK modes.
The ADD model has been studied extensively both theoretically as well as
experimentally. Recently, the extra dimensions are being searched for using 
dilepton and diphoton channels at the
LHC, both by ATLAS \cite{Aad:2012bsa,Aad:2012cy} and CMS \cite{Chatrchyan:2012kc,Chatrchyan:2011fq}. These searches has already put
the lower bound on $M_S$ quite high.

\section{Diffractive Processes}
\label{sec:diffrac_process}

\begin{figure}[!h]
\centering
\subfloat{
\begin{tabular}{ccc}
\resizebox{60mm}{!}{\includegraphics{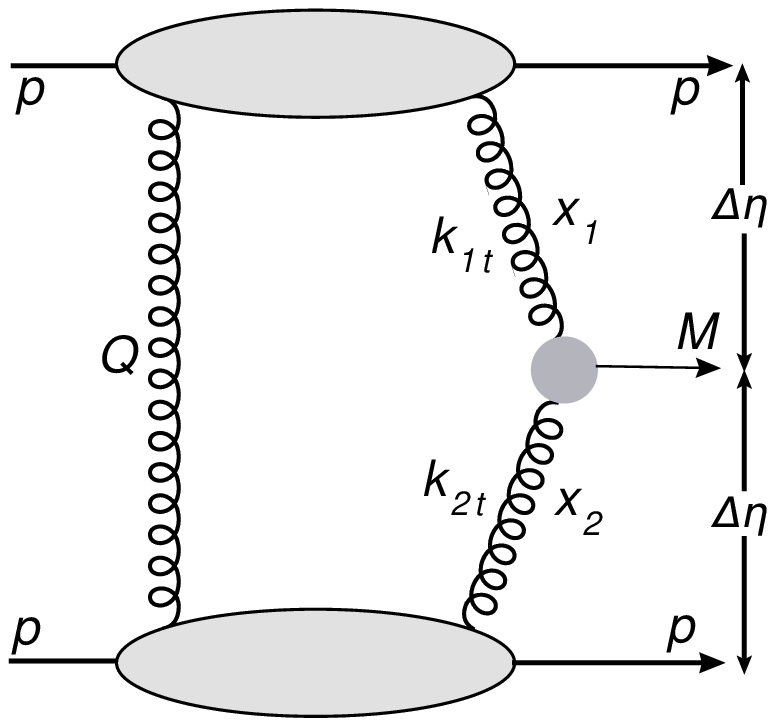}} ~~&&~~
\resizebox{60mm}{!}{\includegraphics{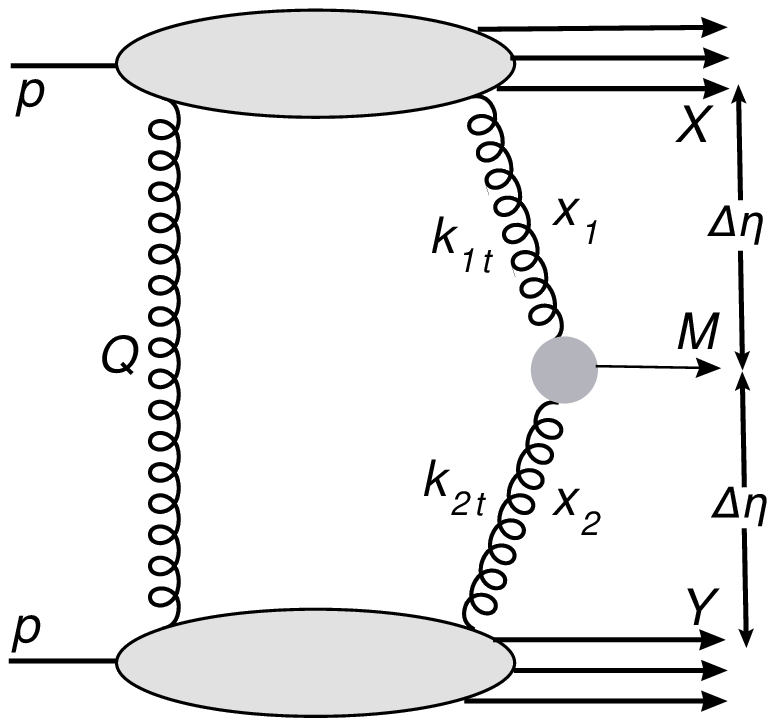}} \\
\hspace{-0.8cm}(a)&&(b)
\end{tabular}}
\caption{\label{fig:DD_process} Double diffractive production of a system of mass 
$M$ with rapidity gap $\Delta\eta$ on either side via (a) the central exclusive process, (b) 
the central inclusive process at the LHC.}
\end{figure}  

We consider two types of double diffractive (DD) mechanisms at the LHC as follows:
\begin{enumerate}
\item 
Central exclusive process (CEP): $pp\to p+M+p$ as shown in 
Fig.~\ref{fig:DD_process}(a) 
\item 
Central inclusive process (CIP): $pp\to X+M+Y$ as shown in 
Fig.~\ref{fig:DD_process}(b) 
\end{enumerate}
Here $M$ denotes the centrally produced system (we also denote the invariant mass 
of the central system with the same notation ``$M$'') and the `$+$' signs denote 
the presence of rapidity gaps between the system $M$ and the outgoing protons (in
exclusive process) or proton dissociated products $X,Y$ (in inclusive process).
The differential cross sections for exclusive and inclusive DD processes shown in
Fig.~\ref{fig:DD_process} can be expressed in the factorized form
~\cite{Khoze:2001xm},
\begin{eqnarray}
M^2\frac{d^2\sigma}{dydM^2} = \mathcal{L}_{gg}(M^2,y)\hat{\sigma}(M^2),
\end{eqnarray}
where $\hat{\sigma}$ is the cross section of the hard subprocess which produces 
the color singlet system of invariant mass $M$ ($gg\to M$) and $\mathcal{L}_{gg}$ 
is the effective luminosity for the production of a system of invariant mass $M$ 
at rapidity $y$. The total cross section $\sigma(pp\to X+M+Y)$ can be calculated using
\begin{equation}
\sigma = \int dydM^2\frac{1}{M^2}\mathcal{L}_{gg}(M^2,y)\hat{\sigma}(M^2)
\end{equation}

In our calculation, we keep $\mathcal{L}_{gg}$ and $\hat{\sigma}$
in a factorized form as the effective luminosity is independent of the subprocess
and for different processes we need to compute just the subprocess cross sections.
Although we keep $\mathcal{L}_{gg}$ and $\hat{\sigma}$ in a factorized form, we 
must remember that there might exist specific selection rules for different
configurations. In the exclusive configuration, the fusing $gg$ state obeys a 
special selection rule in the limit that the protons scatter through zero angle. 
The $z$-component of the angular momentum of a centrally produced system vanishes 
as the incoming state which consists of the fusing hard gluons has $J_z = 0$ with
positive $C$ and $P$ parity~\cite{Khoze:2001xm}. For inclusive
production on the other hand, there is no such selection rule and therefore, the
production rate is larger than that for the exclusive production. Although the
signals are very clear for CEP, the luminosity and the event rates are very small 
due to the restricted kinematics. A much larger phase space is available due to 
the dissociation of the incoming protons in the inclusive DD case.
The luminosities for exclusive and inclusive processes can be computed using
the known parton distribution functions (PDF) of the protons.

\subsection{The Durham Model}

To estimate $\mathcal{L}_{gg}$ we shall use what is known as the ``Durham Model'', 
described in Refs.~\cite{Khoze:1997dr,Khoze:2000cy,Khoze:2001xm,Forshaw:2005qp}. 
The $\mathcal{L}_{gg}$ depends on off-diagonal or skewed gluon distribution 
functions (off-diagonal, since we need to consider the coupling of two gluons with 
different momentum fractions, $x_i$ and $x^{\prime}_i$, with a proton). However, 
as we see, for $x^{\prime}_i\ll x_i$ and $k_{it} \approx Q_{t}$, the relevant
kinematic region for CEP, it is possible to approximate the off-diagonal
distributions using the integrated gluon distribution function
\cite{Shuvaev:1999ce}. For CIP the dominant (leading logarithm) contribution comes
from the asymmetric configuration, $Q_t^2\ll k_{it}^2$. At the partonic level,
the inclusive production is equivalent to the exclusive production and the
unintegrated gluon distributions in partons may be calculated perturbatively in 
terms of non-forward BFKL amplitudes. The effective luminosity depends also on 
two survival probabilities of the rapidity gaps - one to the QCD radiation and
another to the non-perturbative soft rescattering of the protons. The first is
expressed as a Sudakov factor which ensures the fusing gluons remain intact up to 
the hard scale $M/2$. The second factor is included as an explicit multiplicative
factor $\mathcal{S}^2$, known as the ``gap survival factor''. It accounts for the
probability that apart from the perturbative processes, no other particle is 
produced via soft interactions between the protons~\cite{Bjorken:1992er,Gotsman:1993vd,Khoze:2000wk,Kaidalov:2001iz}. In general $\mathcal{S}^2$ is expected to depend on the kinematics of
the process and can be calculated using various QCD based models
\cite{Khoze:2000wk,Kaidalov:2001iz,Khoze:2006uj,Ryskin:2007qx,Gotsman:2008tr,
Ryskin:2009tk}. 
Typically, for LHC energies, $\mathcal{S}^2$ for CEP is estimated to be a few percent~\cite{Khoze:2000wk,Kaidalov:2001iz} (e.g., in \cite{Khoze:2001xm}
authors found $\mathcal{S}^2=0.020~(0.026)$ for 14 TeV (8 TeV) LHC). For CIP, relatively smaller
absorption cross section leads to a larger $\mathcal{S}^2$. 
For simplicity, in our analysis we keep $\mathcal{S}^2=0.1$ for CIP, as a constant multiplicative factor.

We use the MSTW2008 LO parton density~\cite{Martin:2009bu} in our computation. 
Since MSTW parametrization for the gluon distribution function $G(x,Q) = xg(x,Q)$
does not go below $Q = 1$ GeV, for $Q \leq 1$ GeV, we take
the following extrapolation function,
\begin{eqnarray}
G(x,Q) = \alpha(x) Q^{[2+\{\beta(x)-2\}Q]}
\end{eqnarray}
so that for $Q^2 \to 0$, $G(x,Q) \sim Q^2$ \cite{Forshaw:2005qp}. Here $\alpha$
and $\beta$ are the two functions of $x$ and can be computed using the following
relations
\begin{eqnarray}
\beta(x) = \frac{1}{(1+\ln Q_0)}\left(\frac{G^{\prime}(x,Q_0)}{G(x,Q_0)} - \frac{2}{Q_0}\right) + 2;~~\alpha(x) = \frac{G(x,Q_0)}{Q_0^{[2+\{\beta(x) -2\}Q_0]}}
\end{eqnarray}
where $G^{\prime}=\partial G/\partial Q$ and $G,G^{\prime}$ are evaluated at $Q_0=1$ GeV.
To compute 
the Sudakov factor, we use the LO $\alpha_S$ (strong coupling) with $\Lambda_{QCD} = 220$ MeV and $N_f 
=5$. The full 
formalism to calculate effective luminosity for exclusive and inclusive 
configuration is given in Ref.~\cite{Khoze:2001xm}. In Appendix \ref{app:inclu_lum} we briefly
present the computation of $\mathcal{L}_{gg}$ for CIP.


\section{Diffractive Production}
\label{sec:diffrac_produc}
With diffractive processes being dominated by two-gluon exchange as shown in
Fig.~\ref{fig:DD_process}, it is obvious that these processes would be observed 
only for the final states with a substantial coupling to a gluon pair. In this 
work, we consider the central production of KK gravitons ($G$) via DD processes. 
The spin-2 gravitons can have three possible spin projections $J_z=$ 0, 1 and 2. 
Due to the presence of special $J_z=0$ selection rule in the exclusive configuration, 
graviton cannot be produced via leading order (LO) exclusive processes
\footnote{Although there is no $J_z=0$ point-like $gg$ coupling to a $2^+$
graviton, it can still be produced at next to LO (NLO) in the exclusive configuration.} \cite{Khoze:2001xm}.
In the unitary gauge, the coupling of the spin-0 projection 
of $G$ with two gluons is zero as the LO coupling of spin-0 component
with any two vector bosons is proportional to the square of the mass of
the vector boson and gluons are
massless in this case~\cite{Han:1998sg}. By the 
Landau-Yang theorem \cite{Landau:1948kw,Yang:1950rg}, it is impossible to produce a spin-1 particle via the fusion 
of two on-shell vector bosons but the production is not forbidden if those 
vector bosons 
are off-shell. Though the gluons coming from protons are not 
purely on-shell, the degree of virtuality of the gluons is very small. Thus, the 
contribution in 
any process from spin-1 projection of graviton is 
subdominant~\cite{HarlandLang:2010ep}. Thus, the $J_z=2$ is the only possible spin 
projection that can couple to a gluon pair at LO. Therefore, in this paper we
discuss the LHC signatures of $G$ in inclusive configuration.
In what follows we analyze the diphoton and the dilepton channels via the exchange of $G$ arising in ADD type models~\cite{ArkaniHamed:1998rs} as a
signal of extra dimension. We
carefully consider these signal processes and the relevant
background processes for them and look for regions of parameter space where the signal
would dominate. Since these processes can only 
occur in inclusive configuration the event rates could be substantial to search for at the LHC. Moreover, the nature of the intermediate particle and the 
coupling with SM particles, reflects in the angular 
distribution of the final state particles.

\subsection{Dilepton Production}

\begin{figure}[!h]
\centering
\includegraphics[scale=0.8]{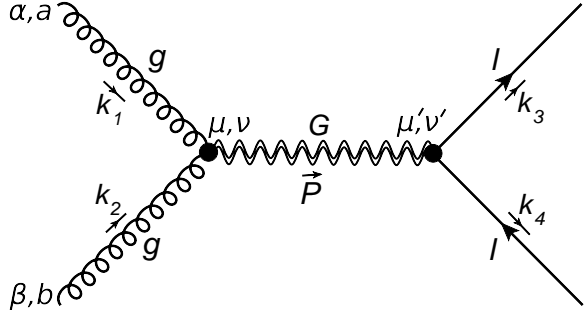}
\caption{\label{fig:dilepton} Subprocess Feynman diagram for dilepton production.}
\end{figure}

We consider the process where dilepton pair is produced from gluon fusion via
the exchange of $G$ as shown in Fig.~\ref{fig:dilepton}.
Feynman rules for this process are taken from Ref.~\cite{Han:1998sg} where one 
can find definitions of all the terms which appear here. We also show these 
Feynman rules in Appendix \ref{app:Feynrules}.
The matrix element for this process is given by
\begin{eqnarray}
\label{eq:M_dilep}
i\mathcal{M}^{ab} = \Big(-\frac{i\kappa}
{2}\delta^{ab}\Big)\epsilon^{\alpha}_1(k_1)\epsilon^{\beta}_2(k_2)V_{\mu\nu;
\alpha\beta}\frac{\frac{i}{2}B^{\mu\nu;\mu^{\prime}\nu^{\prime}}}{P^2-M_{G}^2+i\varepsilon}\Big(
-\frac{i\kappa}{8}\Big)\bar{u}(k_3)\Gamma_{\mu^{\prime}\nu^{\prime}}v(k_4)
\end{eqnarray}
where the coupling $\kappa$ is related to the Newton constant $G_N$ as
$\kappa=\sqrt{16\pi G_N}$. Since the reduced Planck mass $M^{*}_{\textrm{Pl}}=1/\sqrt{8\pi G_N}$ ($M^{*}_{\textrm{Pl}}\approx 2.4\times 10^{18}$ GeV), each KK mode coupling to the SM particles is Planck mass suppressed.
A summation over high multiplicity of KK modes lying below the UV cutoff scale 
$M_S$ compensates the suppression and give rise to a substantial effective coupling strength. 
Therefore, we replace the graviton propagator in Eq. \ref{eq:M_dilep} by the effective propagator 
\begin{equation}
\mathcal{D}_{eff}(\hat{s})=\sum_k \frac{i}{\left[\hat{s} - \left(M_{G}^2\right)_k+i\left(\Gamma_{G}\right)_k 
\left(M_{G}\right)_k\right]}
\end{equation}
where $k$ sums over all KK towers below $M_S$ and $\Gamma_{G}M_{G}=\varepsilon$
and ``hat'' notation is used for subprocess quantities.
Here $\sqrt{\hat s}=M_{ll}$ is the invariant mass of the lepton pair. 
Since, KK modes are quasi-continuous, this summation can be done by defining
KK state density as shown in Appendix \ref{app:Feynrules}~\cite{Han:1998sg}.
For $n$ number of extra dimensions, considering the contributions from 
resonant and nonresonant KK states, the $\mathcal{D}_{eff}$ is given by
\begin{equation}
\label{eq:D_eff}
\mathcal{D}_{eff}(\hat s) = \frac{16\pi \hat{s}^{n/2-1}}{\kappa^2\Gamma(n/2)M_S^{n+2}}\left[\pi + 2iI\left(\frac{M_S}{\sqrt{\hat{s}}}\right)\right]
\end{equation}
where the real part comes from the summation over all resonant contributions 
below $M_S$ and the imaginary part is the summed contribution coming from all the 
nonresonant states. 
The definition of the $I$ function can be found in Appendix \ref{app:Feynrules}.
After summing over the spins of the final state leptons the effective squared 
matrix element  takes the form
\begin{eqnarray}
|\mathcal{M}|^2 &=& \Big(\frac{\kappa^4}{1024}\Big)
(\mathcal{D}_{eff})^2(\epsilon_1^{\alpha_1}(k_1)\epsilon_1^{*\alpha_2}(k_1))
(\epsilon_2^{\beta_1}(k_2)\epsilon_2^{*\beta_2}
(k_2))V_{\mu_1\nu_1
;\alpha_1\beta_1}V_{\mu_2\nu_2;\alpha_2\beta_2}\nonumber\\
&\times & B^{\mu_1\nu_1;\mu_1^{\prime}\nu_1^{\prime}}B^{\mu_2\nu_2;\mu_2^{\prime}\nu_2^{\prime}}
{\textrm {tr}}[(\bar u(k_3)\Gamma_{\mu_1^{\prime}\nu_1^{\prime}}v(k_4))(\bar 
v(k_4)\Gamma_{\mu_1^{\prime}\nu_1^{\prime}}^{\dagger}u(k_3))]\delta_{ab}\delta^{ab}
\end{eqnarray}

In the zero mass limit of the produced lepton, the angular dependence of 
$|\mathcal{M}|^2$ takes a very simple form as follows:
\begin{eqnarray}
|\mathcal{M}|^2 = \delta_{ab}\delta^{ab}\frac{\kappa^4}{64}(\mathcal{D}_{eff})^2\hat s^4(1-\cos^4\theta) 
\end{eqnarray}
where $\theta$ is the scattering angle of the leptons in the center of mass (CM) frame of the 
$G$. Averaging $|\mathcal{M}|^2$ over eight colors and two polarizations of the initial gluons, one can evaluate the subprocess differential cross section after including the phase space factor as
\begin{eqnarray}
\label{eq:dCSdcos_LL}
\frac{d\hat\sigma(gg\rightarrow l^{+}l^{-})}{d|\cos\theta|}=\left(\frac{1}
{32\pi\hat{s}}\right)\cdot\left(\frac{1}{8^2}\right)\cdot\left(\frac{1}{2^2}\right)\cdot 8\cdot
\frac{\kappa^4}{64}(\mathcal{D}_{eff})^2\hat s^4(1-\cos^4\theta) 
\end{eqnarray}

If we consider two types of leptons (i.e. electron and muon) are contributing 
in the dilepton final state, we should multiply the above differential cross section 
by an extra factor of two.

\subsection{Diphoton Production}
\label{sebsec:diphoton}

\begin{figure}[!h]
\centering
\includegraphics[scale=0.8]{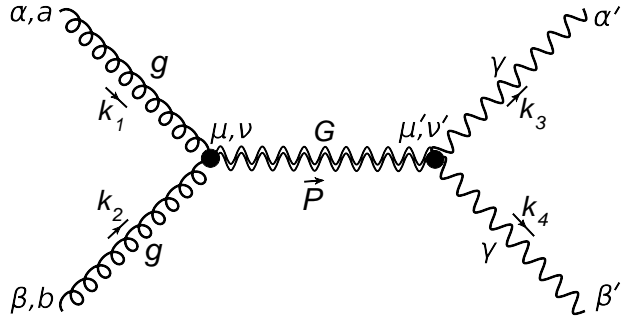}
\caption{\label{fig:diphoton} Subprocess Feynman diagram for diphoton production.}
\end{figure}  

We consider the process where a photon pair is produced from gluon fusion via
the exchange of $G$ as shown in Fig.~\ref{fig:diphoton}. As mentioned earlier,
Feynman rules for this process are taken from Ref.~\cite{Han:1998sg} where one 
can find definitions of all the terms which appear here and also can be found in Appendix \ref{app:Feynrules}. The matrix element for this process is given by
\begin{eqnarray}
i\mathcal{M}^{ab} = \Big(-i\frac{\kappa}
{2}\delta^{ab}\Big)\epsilon^{\alpha}_1(k_1)\epsilon^{\beta}_2(k_2)V_{\mu\nu;
\alpha\beta}\frac{\frac{i}{2}B^{\mu\nu;\mu^{\prime}\nu^{\prime}}}{P^2-M_{G}^2+i\varepsilon} \Big(
-i\frac{\kappa}{2}\Big)V_{\mu^{\prime}\nu^{\prime};
\alpha^{\prime}\beta^{\prime}}\epsilon^{*\alpha^{\prime}}_3(k_3)
\epsilon^{*\beta^{\prime}}_4(k_4)
\end{eqnarray}
After summing over all KK states up to $M_S$ contributing to a physical process
and summing over two polarizations of the final state photons, the effective
squared matrix element takes the form
\begin{eqnarray}
|\mathcal{M}|^2 &=& \Big(\frac{\kappa^4}{64}\Big)
(\mathcal{D}_{eff})^2(\epsilon_1^{\alpha_1}(k_1)\epsilon_1^{*\alpha_2}(k_1))
(\epsilon_2^{\beta_1}(k_2)\epsilon_2^{*\beta_2}(k_2)
)(\epsilon_3^{*\alpha_1^{\prime}}(k_3)\epsilon_3^{\alpha_2^{\prime}}(k_3))
(\epsilon_4^{*\beta_1^{\prime}}(k_4)\epsilon_4^{\beta_2^{\prime}}(k_4))\nonumber\\
&\times & V_{\mu_1\nu_1;\alpha_1\beta_1}V_{\mu_1^{\prime}\nu_1^{\prime};
\alpha_1^{\prime}\beta_1^{\prime}}V_{\mu_2\nu_2;\alpha_2\beta_2}V_{\mu_2^{\prime}\nu_2^{\prime};
\alpha_2^{\prime}\beta_2^{\prime}}B^{\mu_1\nu_1
;\mu_1^{\prime}\nu_1^{\prime}}B^{\mu_2\nu_2;\mu_2^{\prime}\nu_2^{\prime}}\delta_{ab}
\delta^{ab}
\end{eqnarray}
After doing some tedious algebra, the angular dependence of $|\mathcal{M}|^2$ takes the form
\begin{eqnarray}
|\mathcal{M}|^2 = \delta_{ab}\delta^{ab}\frac{\kappa^4}{64}(\mathcal{D}_{eff})^2\hat s^4(1 + 6\cos^2\theta + \cos^4\theta) 
\end{eqnarray}
where $\theta$ is the scattering angle of the photons in the CM frame of the 
$G$. Averaging $|\mathcal{M}|^2$ over eight colors and two polarizations of the initial gluons, one can evaluate the subprocess differential cross section after including the phase space factor as
\begin{eqnarray}
\label{eq:dCSdcos_YY}
\frac{d\hat\sigma(gg\rightarrow \gamma\gamma)}{d|\cos\theta|}=\frac{1}{2}\cdot\left(\frac{1}
{32\pi\hat{s}}\right)\cdot\left(\frac{1}{8^2}\right)\cdot\left(\frac{1}{2^2}\right)\cdot 8\cdot
\frac{\kappa^4}{64}(\mathcal{D}_{eff})^2\hat s^4(1 + 6\cos^2\theta + \cos^4\theta) 
\end{eqnarray}
Here we have included an extra 1/2 factor because of the presence of two identical
particles in the final state.

\section{SM Backgrounds}
\label{sec:SM_bkgrnd}

In the diffractive configuration, there are possibilities to produce a central
system from $gg$, $\gamma\gamma$, $qq$ or $WW$ fusion processes. In a same
kinematic region, the effective luminosity for the $gg$ fusion ($\mathcal{L}_{gg}$) is the largest at the LHC among
all the alternatives. Whereas, the effective luminosities for $\gamma\gamma$, $qq$ or
$WW$ fusion processes are much smaller compared to $\mathcal{L}_{gg}$. Estimations of
effective luminosities for the $gg$, $\gamma\gamma$ and $WW$ can be found in 
\cite{Khoze:2001xm} and for the $qq$ in \cite{Khoze:2004ak}. While computing the SM
backgrounds for the dilepton and the diphoton signal channels, we can sometime neglect
$\gamma\gamma$, $qq$ and $WW$ initiated background processes due to the small luminosities of the fusing particles.
 
\subsection{Dilepton Backgrounds}

\begin{figure}[!h]
\centering
\subfloat{
\begin{tabular}{ccccccc}
\resizebox{35mm}{!}{\includegraphics{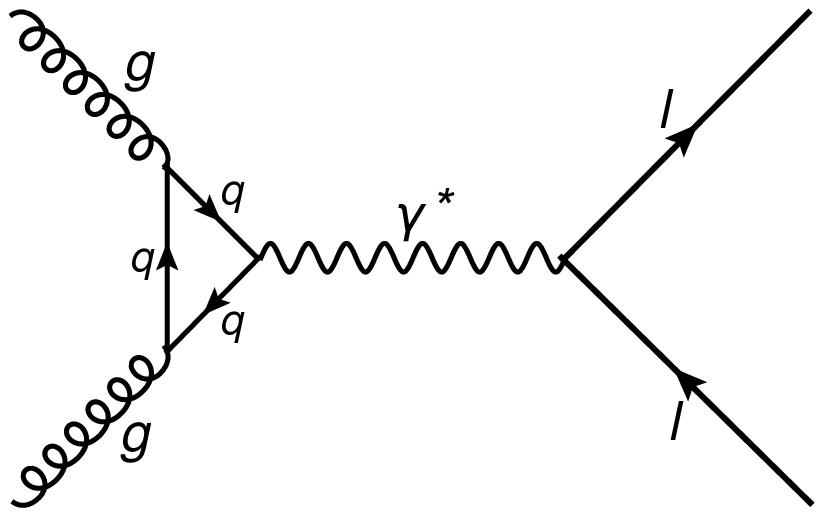}} &&
\resizebox{35mm}{!}{\includegraphics{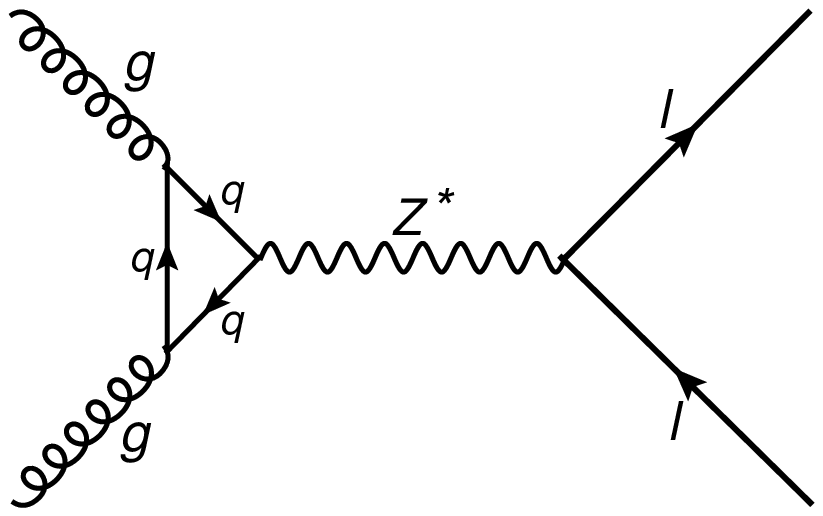}} &&
\resizebox{35mm}{!}{\includegraphics{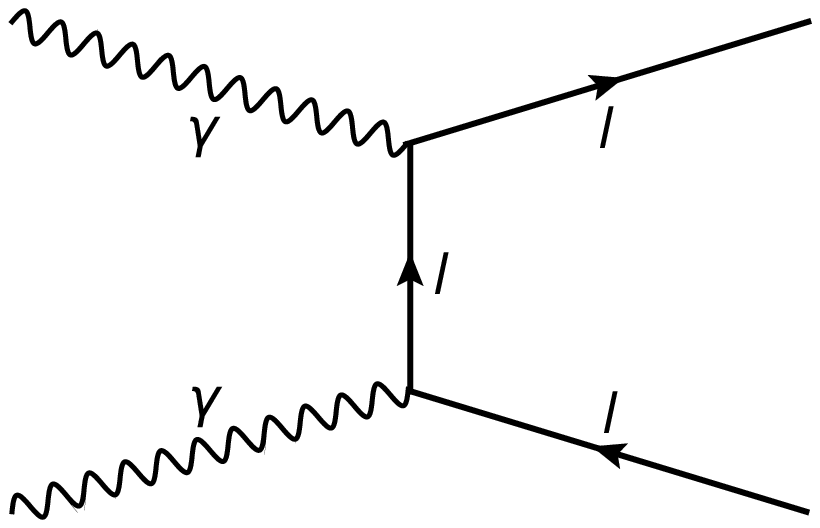}} &&
\resizebox{35mm}{!}{\includegraphics{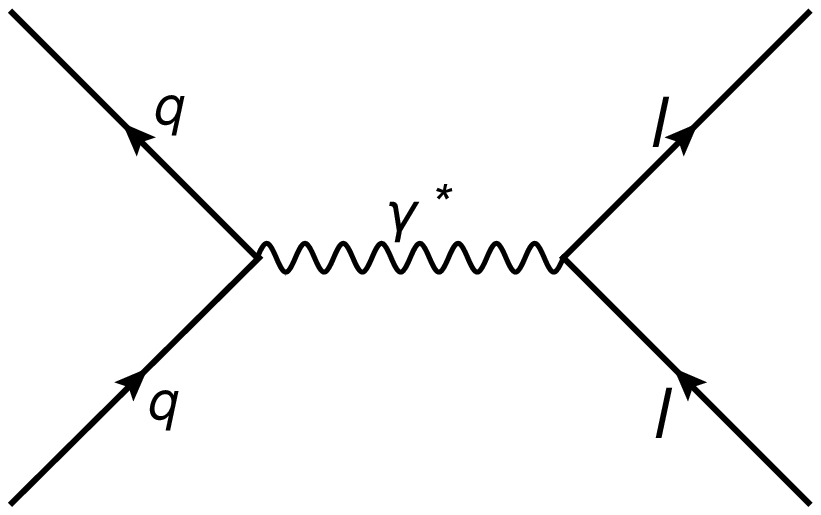}} \\
(a)&&(b)&&(c)&&(d)\\
\resizebox{35mm}{!}{\includegraphics{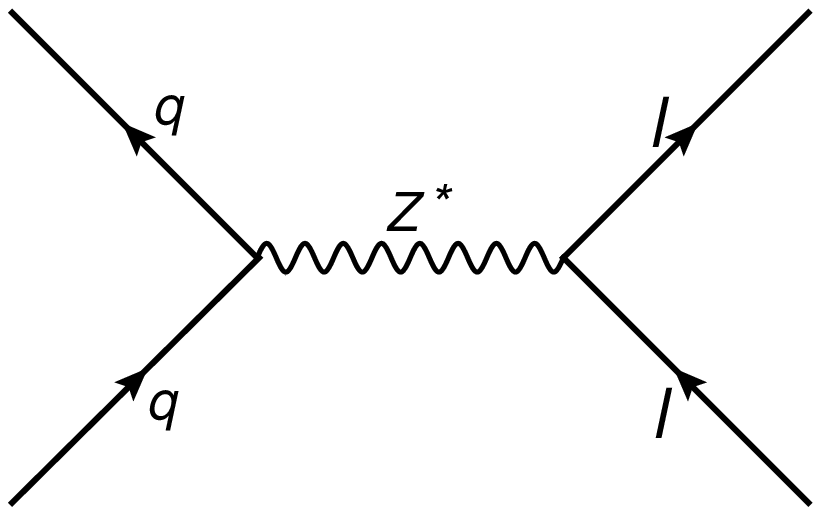}} &&
\resizebox{35mm}{!}{\includegraphics{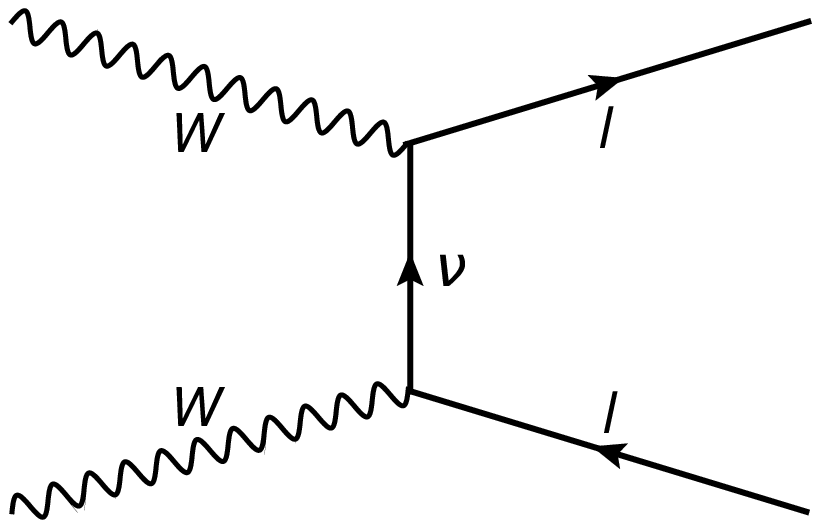}} &&
\resizebox{35mm}{!}{\includegraphics{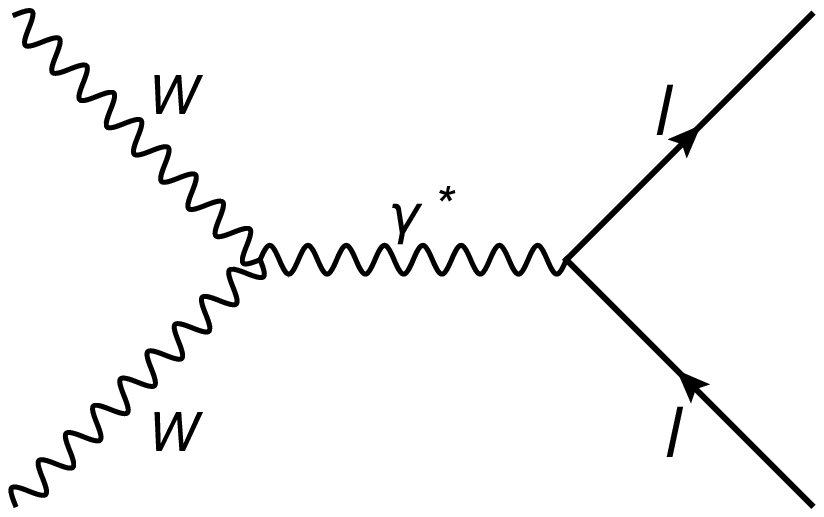}} &&
\resizebox{35mm}{!}{\includegraphics{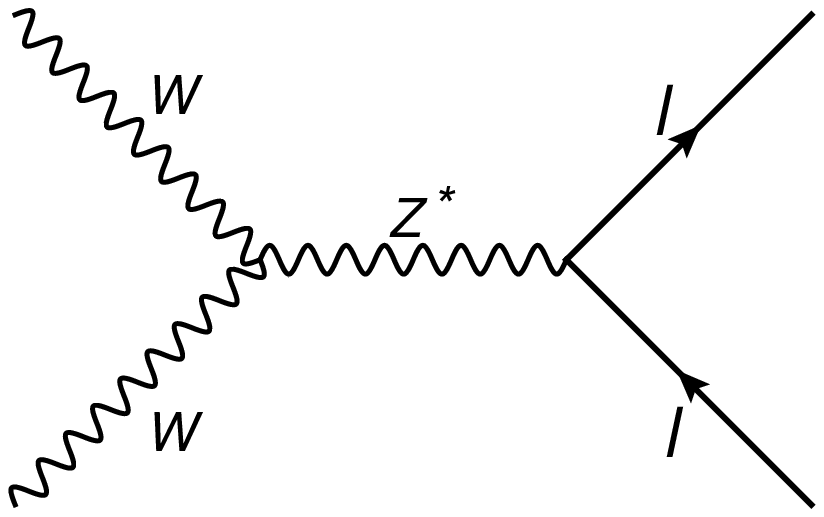}} \\
(e)&&(f)&&(g)&&(h)
\end{tabular}}\\
\caption{\label{fig:dilep_bkgrnd} Subprocess Feynman diagrams for dilepton
channel.}
\end{figure}

In Fig. \ref{fig:dilep_bkgrnd}, we show Feynman diagrams of some important 
background subprocesses for the dilepton channel proceed through $gg$,
$\gamma\gamma$, $qq$ and $WW$ initial states.
When one considers lepton pair production within the SM in the diffractive
configuration, one would not expect a significant rate for a lepton pair 
unaccompanied by other particles (like radiation jets/photons or other SM particles). Although there is always a chance to miss those extra particles other than the
lepton pair or to misidentify other particles as leptons. In our background 
computation we do not include those reducible backgrounds as these 
are expected to be very small. Using Landau-Yang theorem, dual
considerations of angular momentum conservation and Bose symmetry prevents a 
pair of gluons couple to
an on-shell $\gamma/Z$ (via a quark loop, see Figs. \ref{fig:dilep_bkgrnd}(a) and
\ref{fig:dilep_bkgrnd}(b)) or a $WW$ pair couple to an on-shell $\gamma/Z$  
(see Figs. \ref{fig:dilep_bkgrnd}(g) and \ref{fig:dilep_bkgrnd}(h)). While such a coupling, although very small, is indeed permitted for an off-shell $\gamma/Z$. The pieces in the corresponding vertex function are 
proportional to the degree of virtuality of the initial particles or to the
momentum of the $\gamma/Z$ itself. Moreover, $gg\to ll$ processes involves a quark
loop and $WW \to ll$ processes are suppressed due to very low $WW$ luminosity.
Because of these reasons we are not including $gg$ and
$WW$ initiated processes in our computation. We also neglect $qq$ initiated 
processes in Figs. \ref{fig:dilep_bkgrnd}(d) and \ref{fig:dilep_bkgrnd}(e) due to very low luminosity for $qq$.
Consequently, these processes lead only to subdominant contributions.

The dominant SM background for the dilepton channel comes from the inelastic
photo-production of the lepton pair as shown in Fig. \ref{fig:dilep_bkgrnd}(c)
where photons are emitted from the quarks inside protons. This contributions can be very large 
in the kinematic region where the invariant mass of the lepton pair is small. This
is because, the collinear emission of equivalent photons from quark lines can give large contribution. The inelastic $pp\to ll$ cross section has been computed
in Ref. \cite{Drees:1994zx} for Weizs\"acker-Willams photon pairs and an outline
of the computation is presented in Appendix \ref{app:dilep_bkgrnd}. The contributions from the
interference terms among various background processes might be important at
low energies. Since we are only interested in the high energy regions, we do not include them for simplicity.

\subsection{Diphoton Backgrounds}

\begin{figure}[!h]
\centering
\subfloat{
\begin{tabular}{ccccccc}
\resizebox{35mm}{!}{\includegraphics{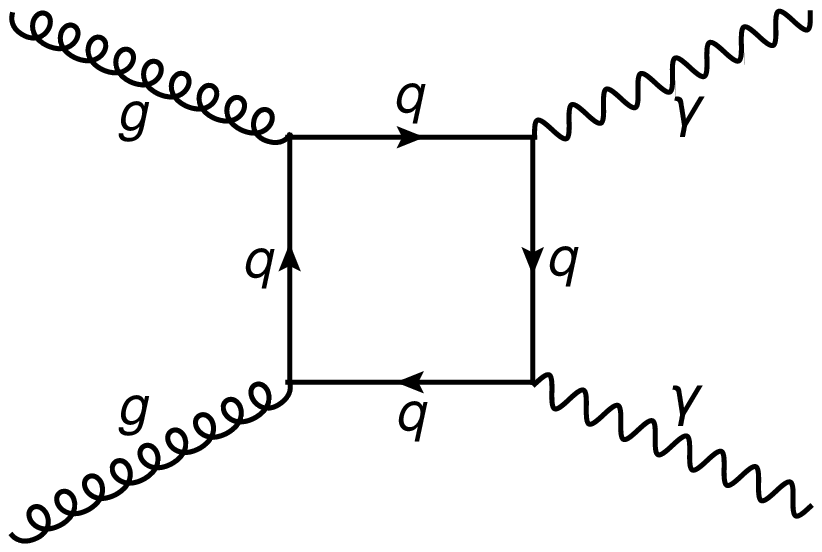}} &&
\resizebox{35mm}{!}{\includegraphics{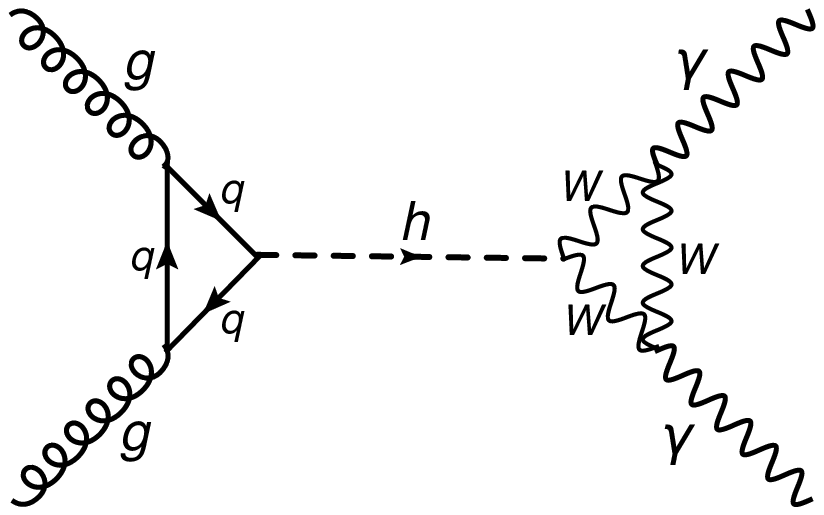}} &&
\resizebox{35mm}{!}{\includegraphics{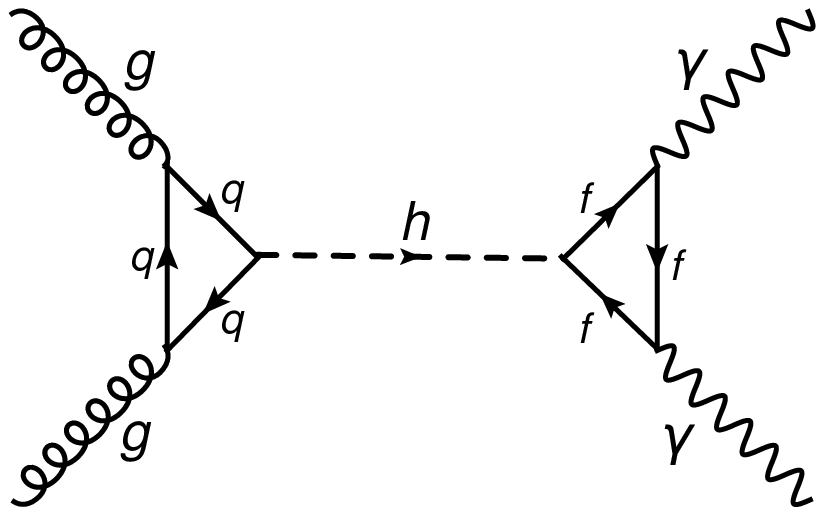}} &&
\resizebox{35mm}{!}{\includegraphics{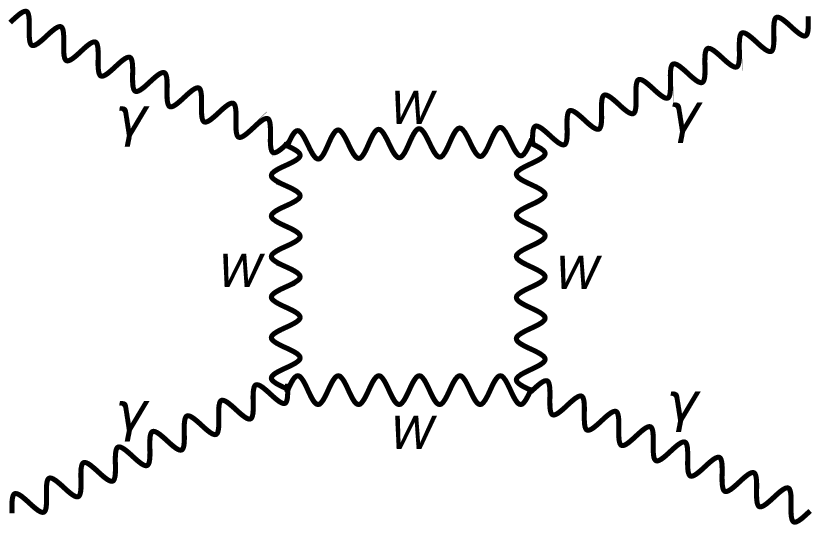}}\\
(a)&&(b)&&(c)&&(d)\\
\resizebox{35mm}{!}{\includegraphics{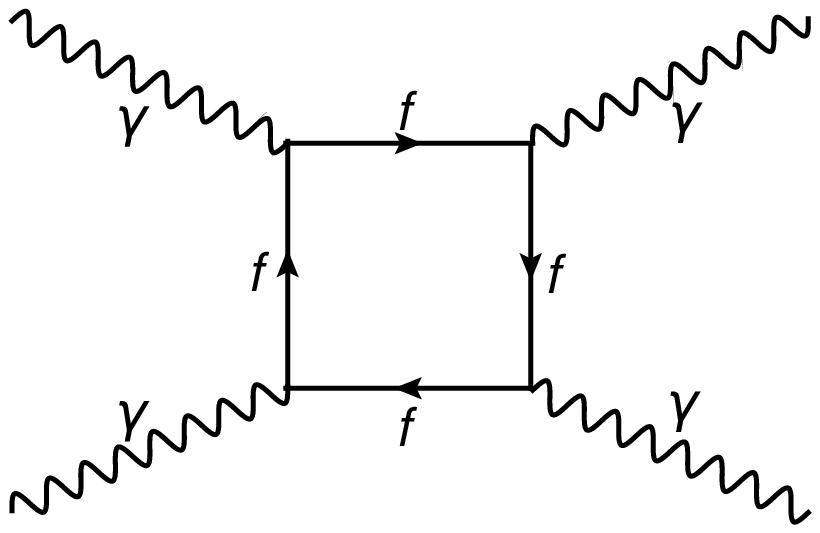}} &&
\resizebox{35mm}{!}{\includegraphics{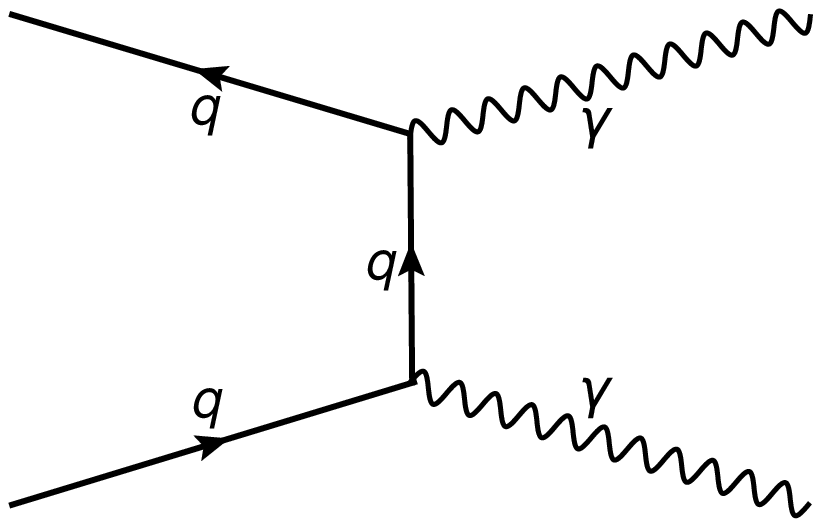}} &&
\resizebox{35mm}{!}{\includegraphics{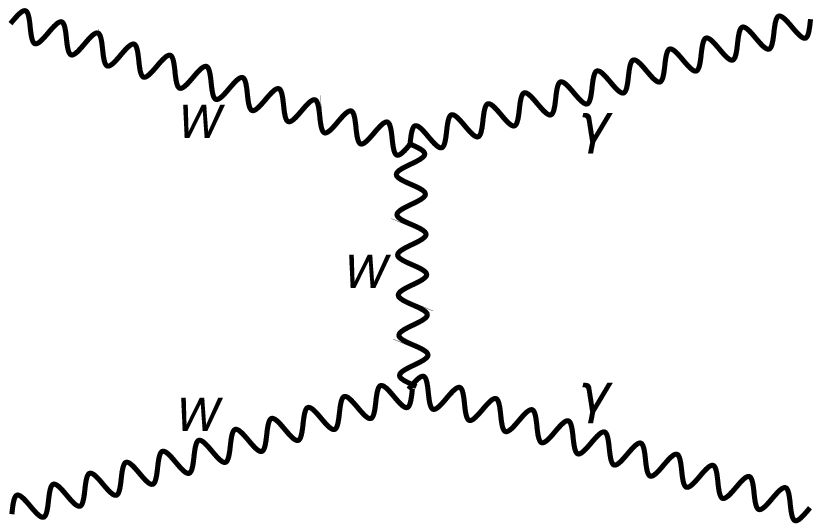}} &&
\resizebox{35mm}{!}{\includegraphics{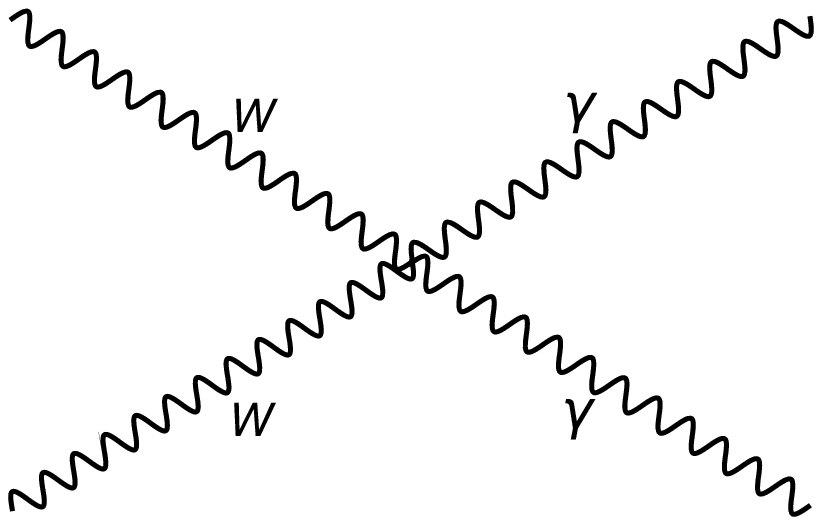}} \\
(e)&&(f)&&(g)&&(h)
\end{tabular}}\\
\caption{\label{fig:dipho_bkgrnd} Subprocess Feynman diagrams for diphoton
channel.}
\end{figure}

In Fig. \ref{fig:dipho_bkgrnd}, we show possible background subprocesses for
the diphoton channel initiated from $gg$, $\gamma\gamma$, $qq$ and $WW$ initial
states. 
The dominant background for the diphoton channel comes from the $gg\to \gamma\gamma$ 
box diagram (via quark loop) shown in Fig. \ref{fig:dipho_bkgrnd}(a)
and cross section is proportional to $\alpha_S^2\alpha_{em}^2$ ($\alpha_{em}$ is the QED coupling). There are other higher order diagrams that contribute to $gg\to \gamma\gamma$ process as shown
in Figs. \ref{fig:dipho_bkgrnd}(b) and \ref{fig:dipho_bkgrnd}(c) where Higgs is involved. The Higgs is produced from $gg$ fusion via a top quark loop and then decayed to $\gamma\gamma$
via another top quark or a $W$ loop. There are also Delbruck type (light-by-light scattering)
backgrounds possible shown in Figs. \ref{fig:dipho_bkgrnd}(d) and \ref{fig:dipho_bkgrnd}(e). In this case quarks from
the dissociated protons, radiate two photons which then give two final state
photons via a fermion or $W$ in the loop. This contribution is
much suppressed as cross section is proportional to $\alpha_{em}^4$. The
other backgrounds which are initiated from $qq$ (see Fig. \ref{fig:dipho_bkgrnd}(f)) and $WW$ (see Figs. \ref{fig:dipho_bkgrnd}(g) and \ref{fig:dipho_bkgrnd}(h)) initial states are negligible as the corresponding effective luminosities are very low.
We note that all the $gg$ and $\gamma\gamma$ initiated background processes occur
only via loop diagrams and they are usually very small. If the signal rate is
substantial in the diphoton channel, searching the signatures of 
extra dimensions in the diphoton channel might be very promising.

We estimate the dominant $gg\to \gamma\gamma$ box diagram background cross section
using $\gamma\gamma \to \gamma\gamma$ helicity amplitudes taken from
\cite{Jikia:1993tc}. Relevant formulae are given in Appendix \ref{app:bkgrnd_YY}. At low energy,
the fermion loop contribution in the box diagram dominates, but above a
few hundred GeV, this contribution starts falling rapidly. Whereas, the $W$ loop
contribution dominates in the total cross section of the $\gamma\gamma \to \gamma\gamma$ process at high energies. Other higher order $\gamma\gamma \to h\to \gamma\gamma$ processes similar to Figs. \ref{fig:dipho_bkgrnd}(b) and \ref{fig:dipho_bkgrnd}(c) are not considered here. We do not include
$qq$ initiated process in our background computation as the contribution
originates from $q\bar{q}$ $t$-channel exchange in Fig. \ref{fig:dipho_bkgrnd}(f) is roughly two order of magnitude lower
than the $gg$ contributions \cite{Khoze:2004ak}. As mentioned earlier, we neglect
$WW$ initiated processes for low $WW$ luminosity and all background interference terms for simplicity.

\section{Numerical Results}
\label{sec:numerical_result}

In all our numerical computations we use the MSTW2008 LO parton densities to obtain the 
unintegrated parton distributions. Here we once again give the numerical values
of some quantities which are used. To compute the Sudakov form factors, we use NLO $\alpha_S$ with $\Lambda_{QCD}=220$ MeV and number of ``active'' flavor $N_f=5$. We take a fixed
value of the ``gap survival factor'', $\mathcal{S}^2=0.1$ in our analysis. 
Our results depend on two model parameters, the number of extra dimensions $n$
and the ultra violet cutoff $M_S$ of the theory and four
kinematical variables, invariant mass $M$ and rapidity $y$ of the centrally produced system, 
rapidity gap $\Delta\eta$ between the central system and the outgoing
protons or proton dissociated products, and the CM energy $\sqrt{S}$ of
two colliding protons at the LHC. The cutoff scale $M_S$ is of the order of 
few TeV for
low scale gravity theories. The lower limit of $M_S$ is bounded 
from experiments and the bounds are already quite high as stated in Sec. \ref{sec:ADDintro}. If we increase $M_S$, the signal cross sections decrease
as usual. In our case the signal falls off very rapidly with increasing 
$M_S$ because of the presence of $M_S^8$ in the denominator of the signal 
cross sections (see Eqs. \ref{eq:dCSdcos_LL} and \ref{eq:dCSdcos_YY}). This makes the ADD model very difficult to 
search for experimentally. 

\begin{figure}[!h]
\centering
\subfloat{
\begin{tabular}{ccc}
\resizebox{80mm}{!}{\includegraphics{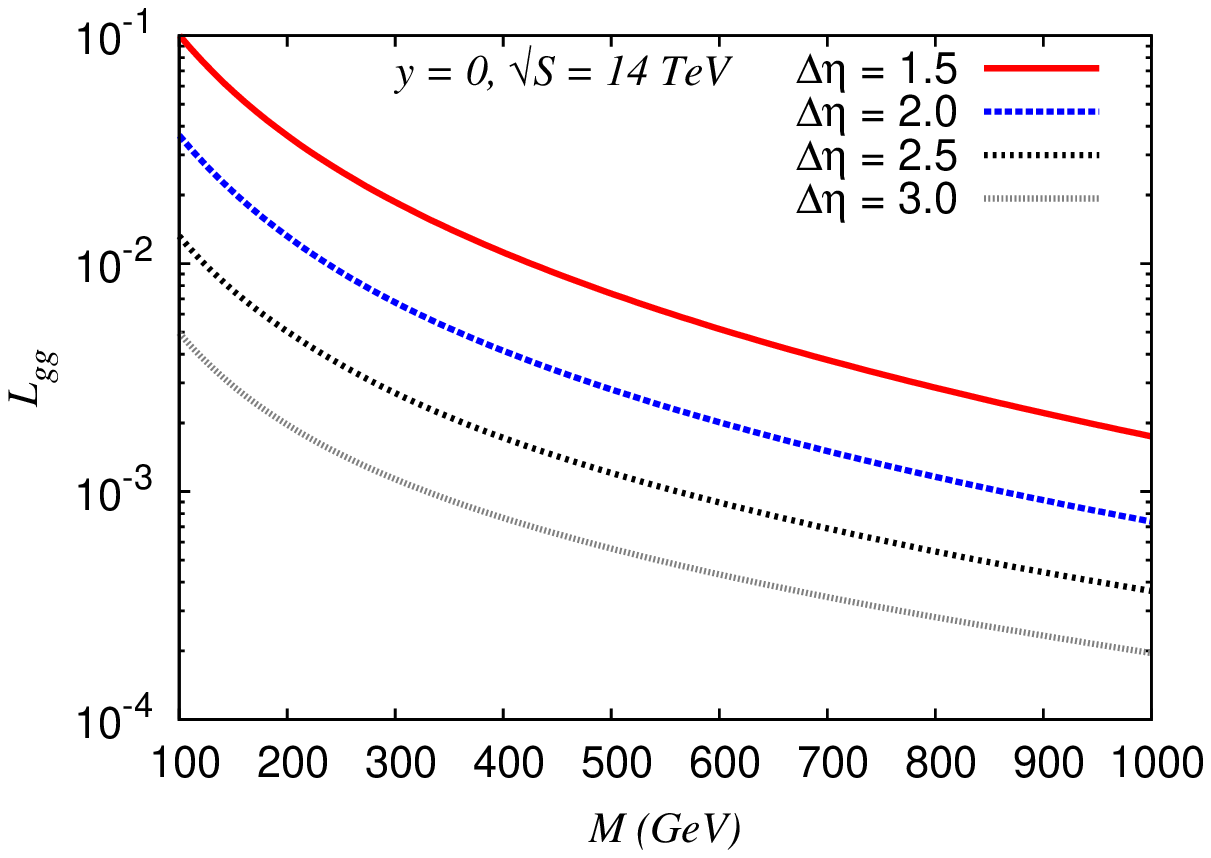}} &&
\resizebox{80mm}{!}{\includegraphics{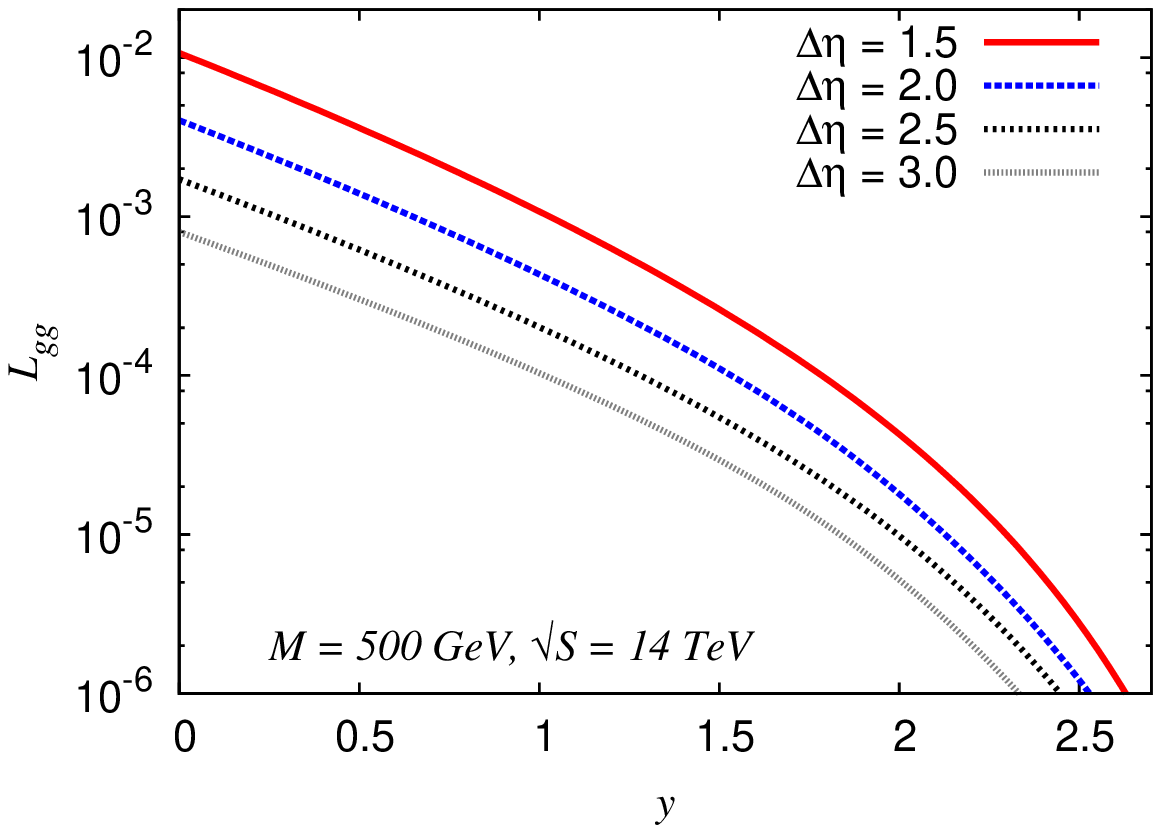}} \\
\footnotesize{\hspace{1.1cm}(a)}&&\footnotesize{\hspace{1.2cm}(b)}
\end{tabular}}
\caption{\label{fig:luminosity} (a) The effective luminosity $\mathcal{L}_{gg}$ 
as a function of the invariant mass $M$ of the central system produced at $y=0$ at the
14 TeV LHC and (b) $\mathcal{L}_{gg}$ as a function of $y$ for the production of a
system with mass $M=500$ GeV at the 14 TeV LHC. For both the plots we choose
four values of the rapidity gap, i.e. $\Delta\eta=1.5$ 2.0, 2.5 and 3.0.}
\end{figure}

In Fig. \ref{fig:luminosity}(a) we show the dependence of the effective luminosity $\mathcal{L}_{gg}$ on the invariant mass $M$ of the central system produced
at rapidity region $y=0$ at the 14 TeV LHC. The $\mathcal{L}_{gg}$ shows an 
decreasing nature with increasing $M$ since it is harder to produce more massive
objects while keeping other parameters fixed. In Fig. \ref{fig:luminosity}(b) we 
plot the dependence of $\mathcal{L}_{gg}$ on $y$ with $M=500$ GeV at the 14 TeV LHC. In the diffractive processes it is more
likely to produce an object centrally i.e. near $y=0$ region. Thus, a system 
which is produced away from the central region is less probable and we see 
a decreasing behavior of $\mathcal{L}_{gg}$ with increasing $y$.

\begin{figure}[!h]
\centering
\subfloat{
\begin{tabular}{ccc}
\resizebox{80mm}{!}{\includegraphics{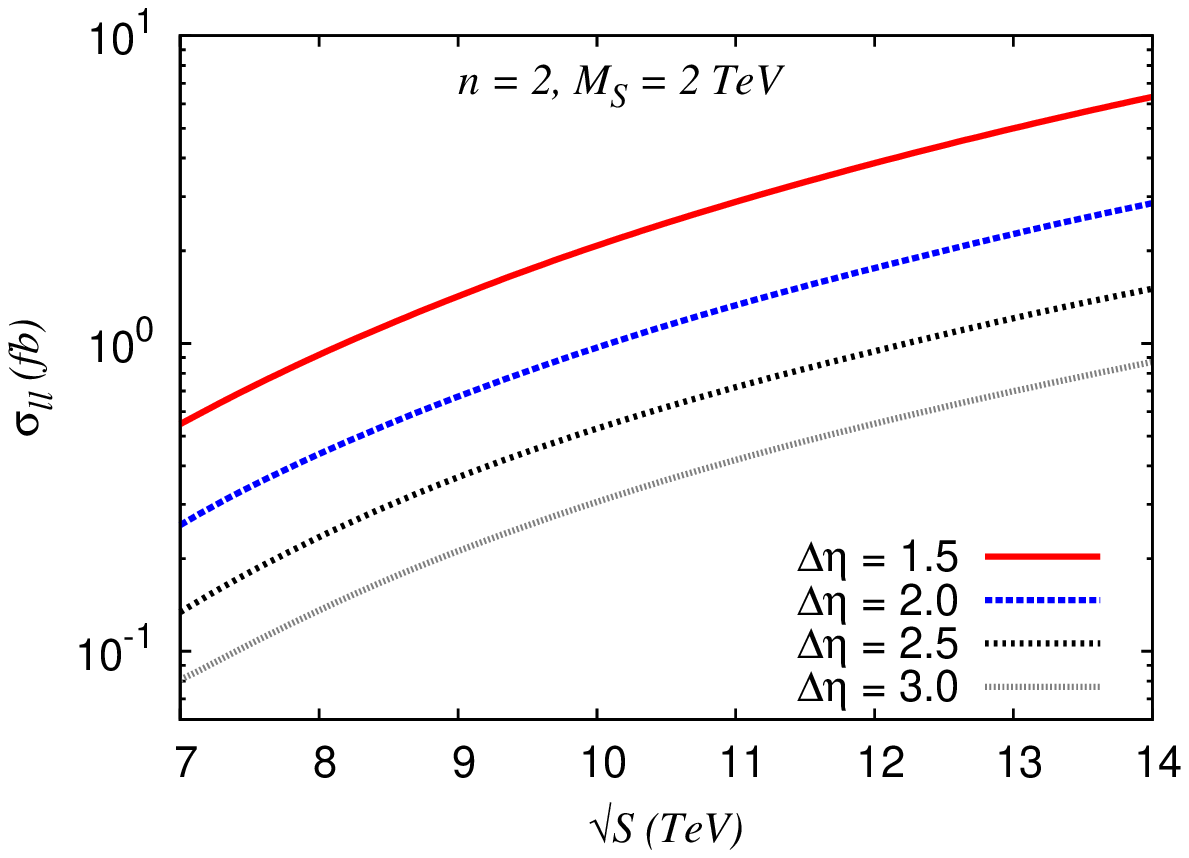}} &&
\resizebox{80mm}{!}{\includegraphics{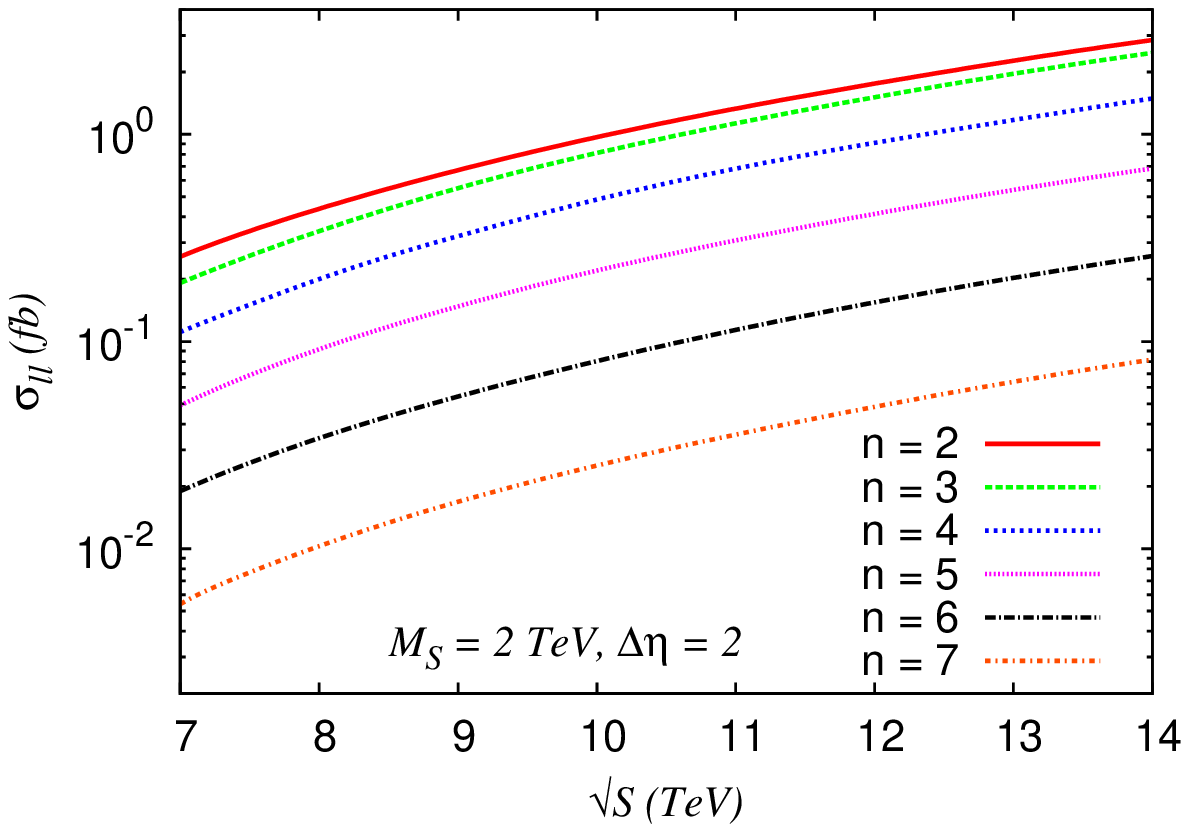}} \\
\footnotesize{\hspace{1.1cm}(a)}&&\footnotesize{\hspace{1.2cm}(b)}
\end{tabular}}
\caption{\label{fig:tCS_sqrtS_LL} The total signal cross section ($pp\to X+ll+Y$) in dilepton
channel ($\sigma_{ll}$) as a function of LHC CM energy. In (a) we plot $\sigma_{ll}$ for $\Delta\eta=$ 1.5, 2.0, 2.5 and 3.0 for $n=2$ and $M_S=2$ TeV. In (b) we plot $\sigma_{ll}$ for $n=2-7$ for $M_S=2$ TeV and $\Delta\eta=2$.
In both plots $\sigma_{ll}$'s are computed after applying the cut defined in Eq. \ref{eq:tCS_cuts1}.}
\end{figure}

In Fig. \ref{fig:tCS_sqrtS_LL} we show the total signal cross section in the 
dilepton ($\sigma_{ll}$ where $l=\{e,\mu\}$) channel as a function of $\sqrt{S}$.
In Fig. \ref{fig:tCS_sqrtS_LL}(a) we show $\sigma_{ll}$ for different $\Delta\eta$
taking $n=2$ and $M_S=2$ TeV. The cross section increases as we increase 
$\sqrt{S}$ as expected. If we increase $\Delta\eta$, the $\sigma_{ll}$ decreases as the effective luminosity decreases with increasing $\Delta\eta$ (see Figs. 
\ref{fig:luminosity}(a) and \ref{fig:luminosity}(b)).
In Fig. \ref{fig:tCS_sqrtS_LL}(b) we show the $\sigma_{ll}$ for different $n$ 
varied from two to seven. We see that as we increase $n$, the $\sigma_{ll}$ 
decreases. This is because $\kappa^2|\mathcal{D}_{eff}| \sim 2/(n-2)$ for $n>2$,
in the limit $M_S^2\gg \hat{s}$ (see Eq. \ref{eq:apprx_Ds}).
Here we compute the total cross section
after applying the following selection cuts:
\begin{equation}
\label{eq:tCS_cuts1}
|y_{\psi\psi}| \leq 2;~~|\cos\theta_{\psi\psi}|\leq 1;~~100~\textrm{GeV} \leq M_{\psi\psi} < M_S - 10~\textrm{GeV}
\end{equation}
where $\theta$ is the scattering angle of $\psi$ (where $\psi = l,\gamma$) in the 
rest frame of the centrally produced system and $M_{\psi\psi}$ is the invariant
mass of the $\psi\psi$ pair. We have collected events with the
rapidity of the central system $|y_{\psi\psi}| \leq 2$. The upper limit of $M_{\psi\psi}$ can go to $M_S$ up to which the theory is 
valid. Here we have taken the upper limit on $M_{\psi\psi}$ slightly smaller than $M_S$ as
the $I$ function in Eq. \ref{eq:D_eff} diverges at $M_S$. The cross sections are 
only few $fb$ and we need higher CM energy at the LHC to observe these events.
 
\begin{figure}[!h]
\centering
\subfloat{
\begin{tabular}{ccc}
\resizebox{80mm}{!}{\includegraphics{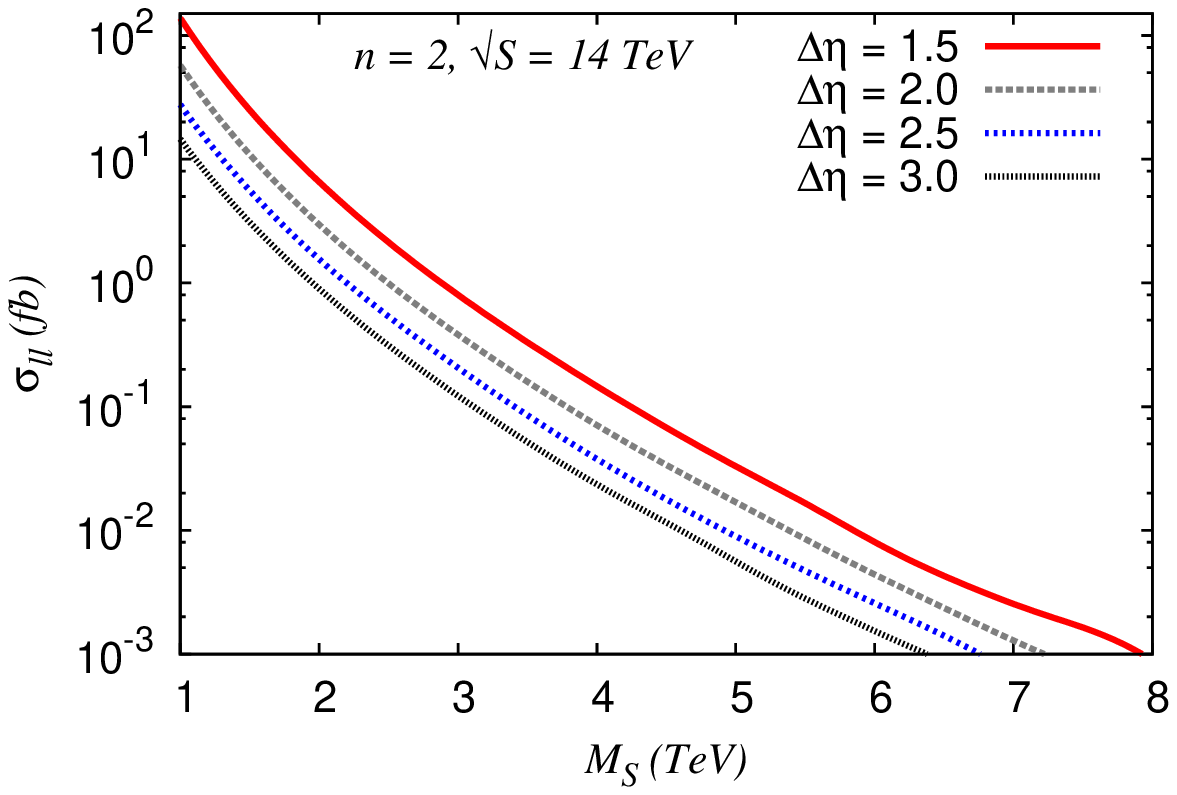}} &&
\resizebox{80mm}{!}{\includegraphics{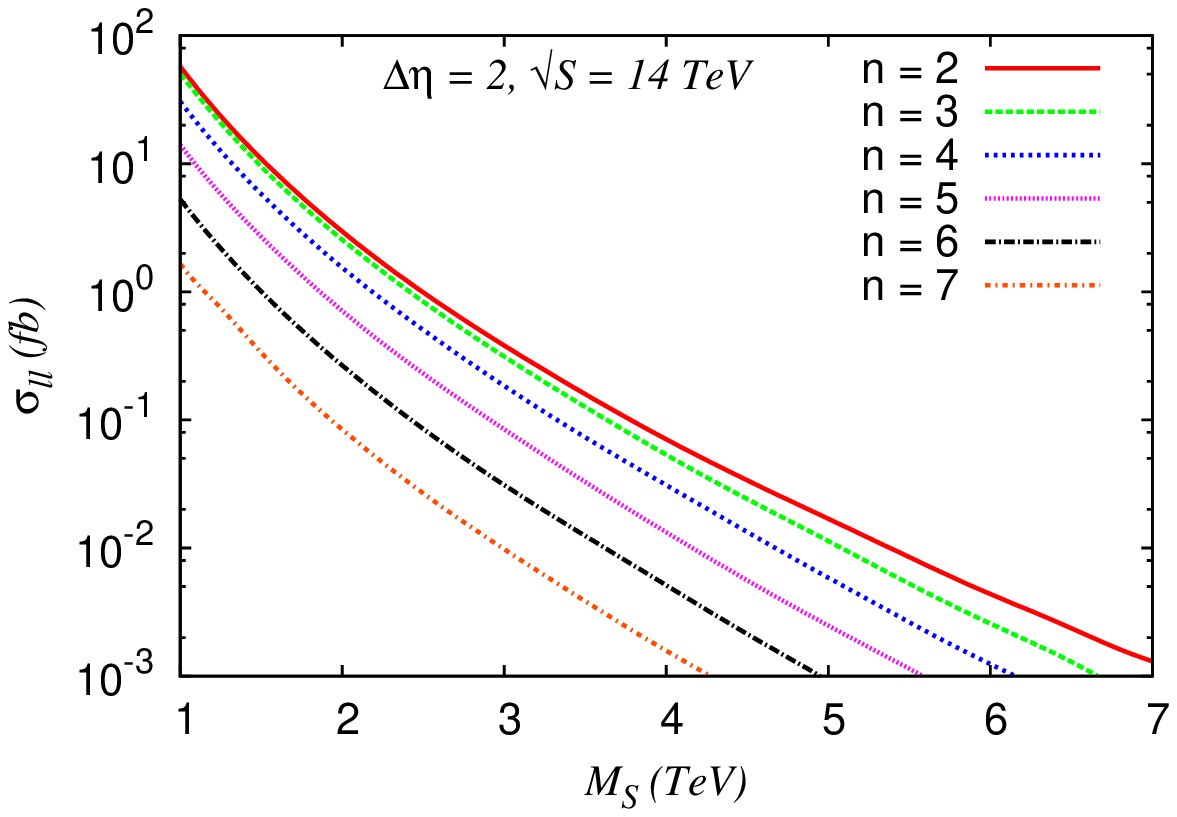}} \\
\footnotesize{\hspace{1.1cm}(a)}&&\footnotesize{\hspace{1.2cm}(b)}
\end{tabular}}
\caption{\label{fig:tCS_Ms_LL} The total signal cross sections ($pp\to X+ll+Y$) in dilepton
channel ($\sigma_{ll}$) as a function of $M_S$ at the 14 TeV LHC. In (a) we plot $\sigma_{ll}$ for $\Delta\eta=$ 1.5, 2.0, 2.5 and 3.0 for $n=2$. In (b) we plot $\sigma_{ll}$ for $n=2-7$ for $\Delta\eta=2$.
In both plots $\sigma_{ll}$'s are computed after applying the cut defined in Eq. \ref{eq:tCS_cuts1}.}
\end{figure}

In Fig. \ref{fig:tCS_Ms_LL} we show the total signal cross section in the dilepton  ($\sigma_{ll}$) channel as a function of $M_S$ at the 14 TeV LHC.
In Fig. \ref{fig:tCS_Ms_LL}(a) we show $\sigma_{ll}$ for different $\Delta\eta$
taking $n=2$ and in Fig. \ref{fig:tCS_Ms_LL}(b) we show $\sigma_{ll}$ for different $n$ taking $\Delta\eta=2$. Here we compute the total cross section
after applying the kinematical cuts defined in Eq. \ref{eq:tCS_cuts1}.
The cross section decreases rapidly as we increase $M_S$ since $M_S^8$ is present in the denominator of the total cross section.

\begin{figure}[!h]
\centering
\subfloat{
\begin{tabular}{ccc}
\resizebox{80mm}{!}{\includegraphics{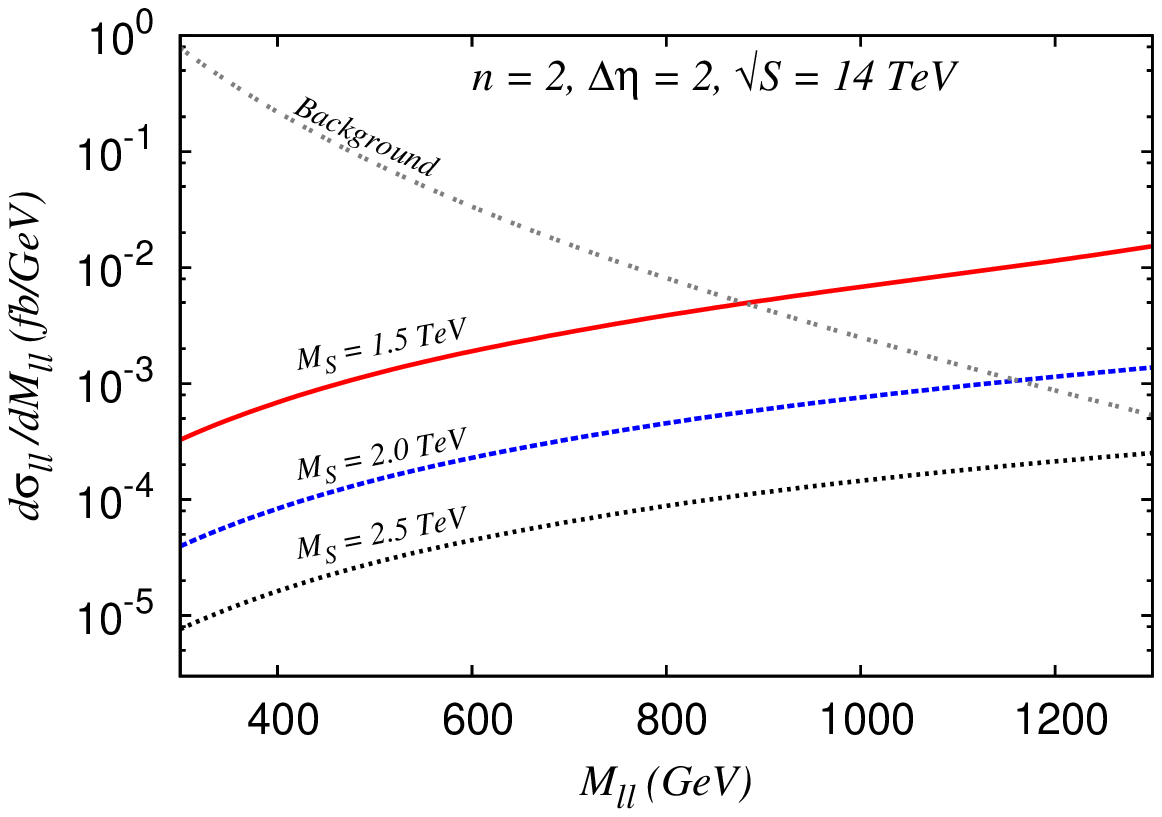}} &&
\resizebox{80mm}{!}{\includegraphics{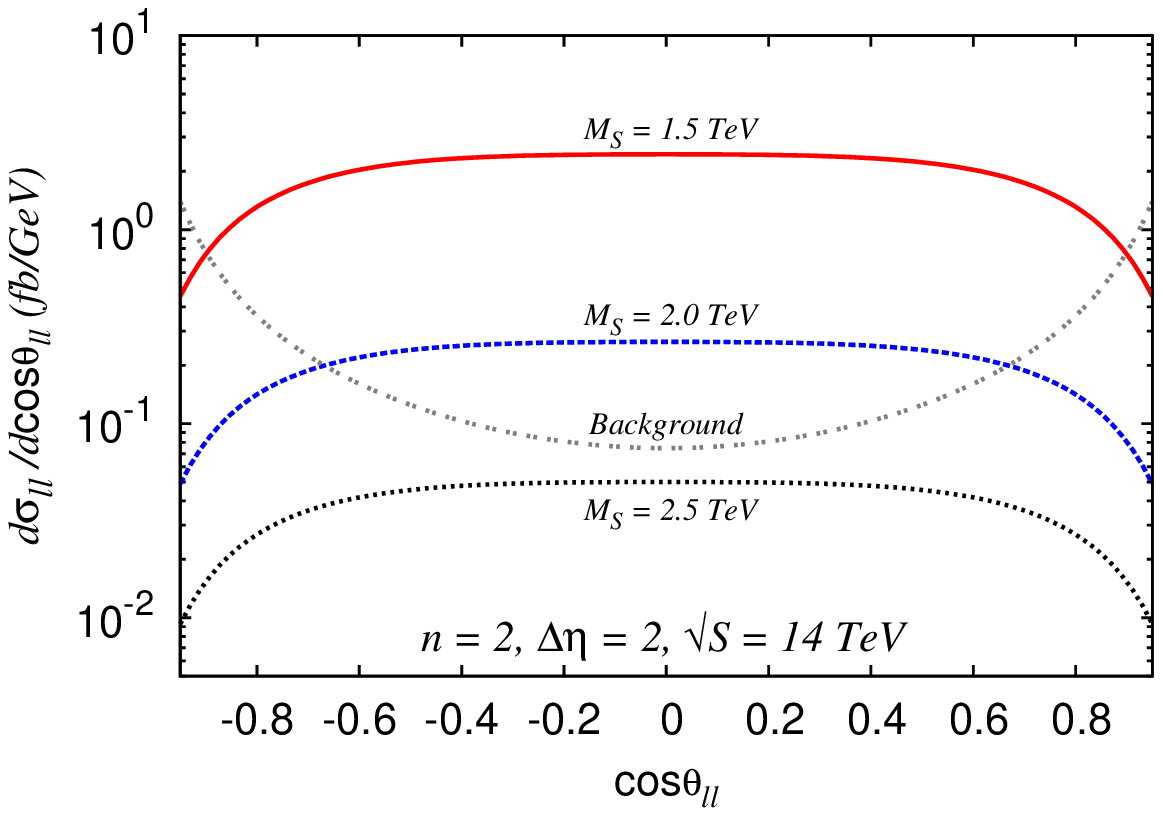}}\\
\footnotesize{\hspace{1.1cm}(a)}&&\footnotesize{\hspace{1.2cm}(b)}
\end{tabular}}
\caption{\label{fig:dCS_LL} (a) The invariant mass distributions ($d\sigma_{ll}/dM_{ll}$) for signal and background ($pp\to X+ll+Y$) of the lepton pair after applying the cuts defined in Eq. \ref{eq:dCS_cuts1} at the 14 TeV LHC. (b) The angular distributions ($d\sigma_{ll}/d\cos\theta_{ll}$) for signal and background of the lepton pair after applying the cuts defined in Eq. \ref{eq:dCS_cuts2} at the 14 TeV LHC. For both plots we take $M_S=$ 1.5, 2.0 and 2.5 TeV and $n=2$ for signal and $\Delta\eta=2$ for both signal and 
background.}
\end{figure}  

In Fig. \ref{fig:dCS_LL}(a) we show signal and background invariant mass 
distributions ($d\sigma_{ll}/dM_{ll}$) of the lepton pair after applying the selection 
cuts defined in Eq. \ref{eq:dCS_cuts1} at the 14 TeV LHC. To compute $d\sigma_{ll}/dM_{ll}$ we have not
integrated $\cos\theta_{\psi\psi}$ in the full range from $-1$ to $1$ because the
background cross section diverges as $|\cos\theta_{\psi\psi}|\to 1$ 
(see Eq. \ref{eq:dCSdcosL_bkgrndLL}).
We plot signal 
$d\sigma_{ll}/dM_{ll}$ for $M_S=$ 1.5, 2.0 and 2.5 TeV taking $n=2$ and, 
$\Delta\eta=2$ for both signal
and background. The signal $d\sigma_{ll}/dM_{ll}$ is a
monotonically increasing function of $M_{ll}$ ($M_{ll}^2=\hat{s}$) because the subprocess cross section
increases roughly as $\hat{s}^3$ (see Eq. \ref{eq:dCSdcos_LL}). The background
cross section is computed only considering the dominant $\gamma\gamma\to ll$
channel. For small values of $M_{ll}$ the background is quite large but falls 
off rapidly as we increase $M_{ll}$. Thus, to observe an excess over the 
background one should collect lepton pairs with sufficiently high $M_{ll}$.
\begin{equation}
\label{eq:dCS_cuts1}
|y_{ll}| \leq 2;~~|\cos\theta_{ll}|\leq 0.8
\end{equation} 

In Fig. \ref{fig:dCS_LL}(b) we show signal and background angular
distributions ($d\sigma_{ll}/d\cos\theta_{ll}$) of the lepton pair after applying the selection cuts defined in Eq. \ref{eq:dCS_cuts2} at the 14 TeV LHC.
To compute $d\sigma_{ll}/d\cos\theta_{ll}$ we apply a very high invariant mass cut
on the lepton pair to reduce large background in the small $M_{ll}$ region. We
show signal $d\sigma_{ll}/d\cos\theta_{ll}$ for $M_S=$ 1.5, 2.0 and 2.5 TeV taking 
$n=2$ and, $\Delta\eta=2$ for both signal and background. The angular distribution carry spin information of the
intermediate particle. For dilepton production, the angular distributions for signal and background
show a very contrasting behavior. For signal $d\sigma_{ll}/d\cos\theta_{ll}\sim 
(1-\cos^4\theta_{ll})$ (see Eq. \ref{eq:dCSdcos_LL}) which attains its maximum at $\cos\theta_{ll}=0$. On the other hand,
background attains its minimum at $\cos\theta_{ll}=0$ since it goes as $(1+\cos^2\theta_{ll})/\sin^2\theta_{ll}$ and diverges when $|\cos\theta_{ll}|\to 1$
(see Eq. \ref{eq:dCSdcosL_bkgrndLL}).
This contrasting feature can be used as a unique signature of the extra
dimensions.
\begin{equation}
\label{eq:dCS_cuts2}
|y_{ll}| \leq 2;~~750 ~\textrm{GeV} \leq M_{ll} < 1250~\textrm{GeV} 
\end{equation}

\begin{figure}[!h]
\centering
\subfloat{
\begin{tabular}{ccc}
\resizebox{80mm}{!}{\includegraphics{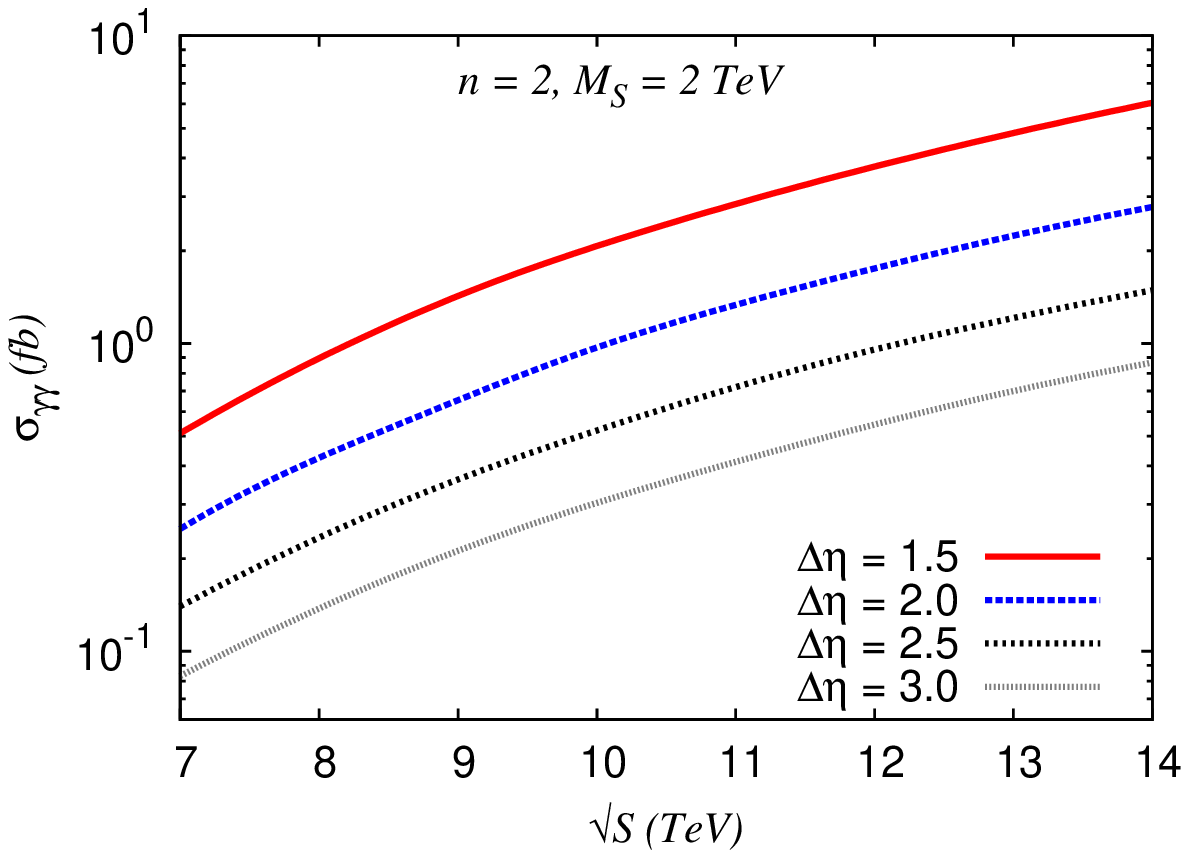}} &&
\resizebox{80mm}{!}{\includegraphics{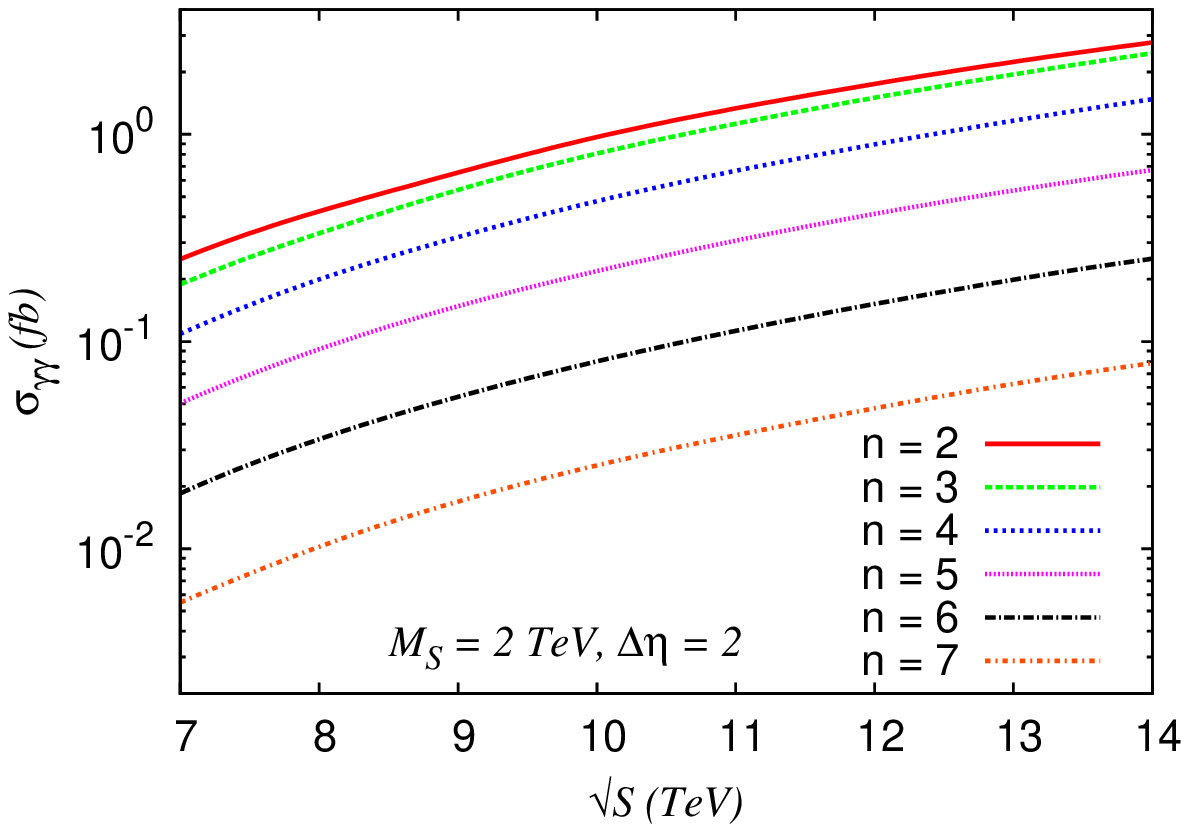}} \\
\footnotesize{\hspace{1.1cm}(a)}&&\footnotesize{\hspace{1.2cm}(b)}
\end{tabular}}
\caption{\label{fig:tCS_sqrtS_YY} The total signal cross sections 
($pp\to X+\gamma\gamma+Y$) in diphoton
channel ($\sigma_{ll}$) as a function of LHC CM energy. In (a) we plot $\sigma_{ll}$ for $\Delta\eta=$ 1.5, 2.0, 2.5 and 3.0 for $n=2$ and $M_S=2$ TeV. In (b) we plot $\sigma_{ll}$ for $n=2-7$ for $M_S=2$ TeV and $\Delta\eta=2$.
In both plots $\sigma_{ll}$'s are computed after applying the cut defined in Eq. \ref{eq:tCS_cuts1}.}
\end{figure}

\begin{figure}[!h]
\centering
\subfloat{
\begin{tabular}{ccc}
\resizebox{80mm}{!}{\includegraphics{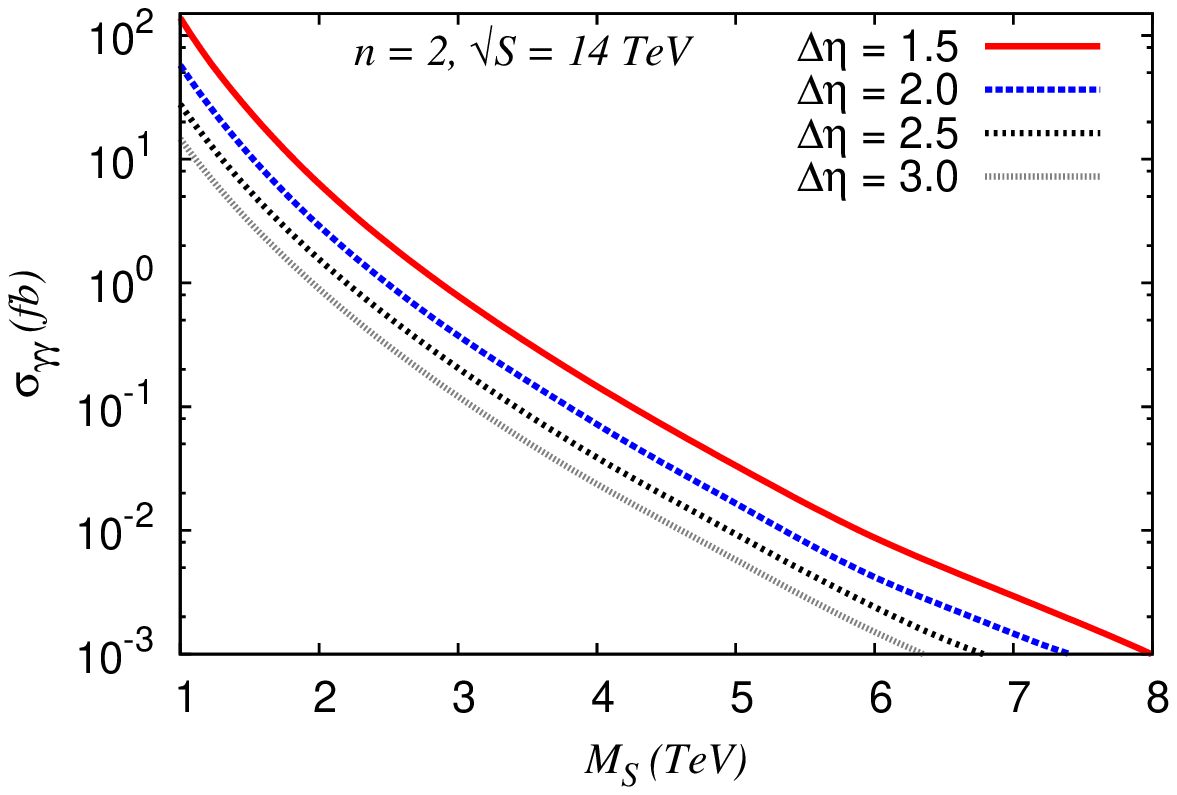}} &&
\resizebox{80mm}{!}{\includegraphics{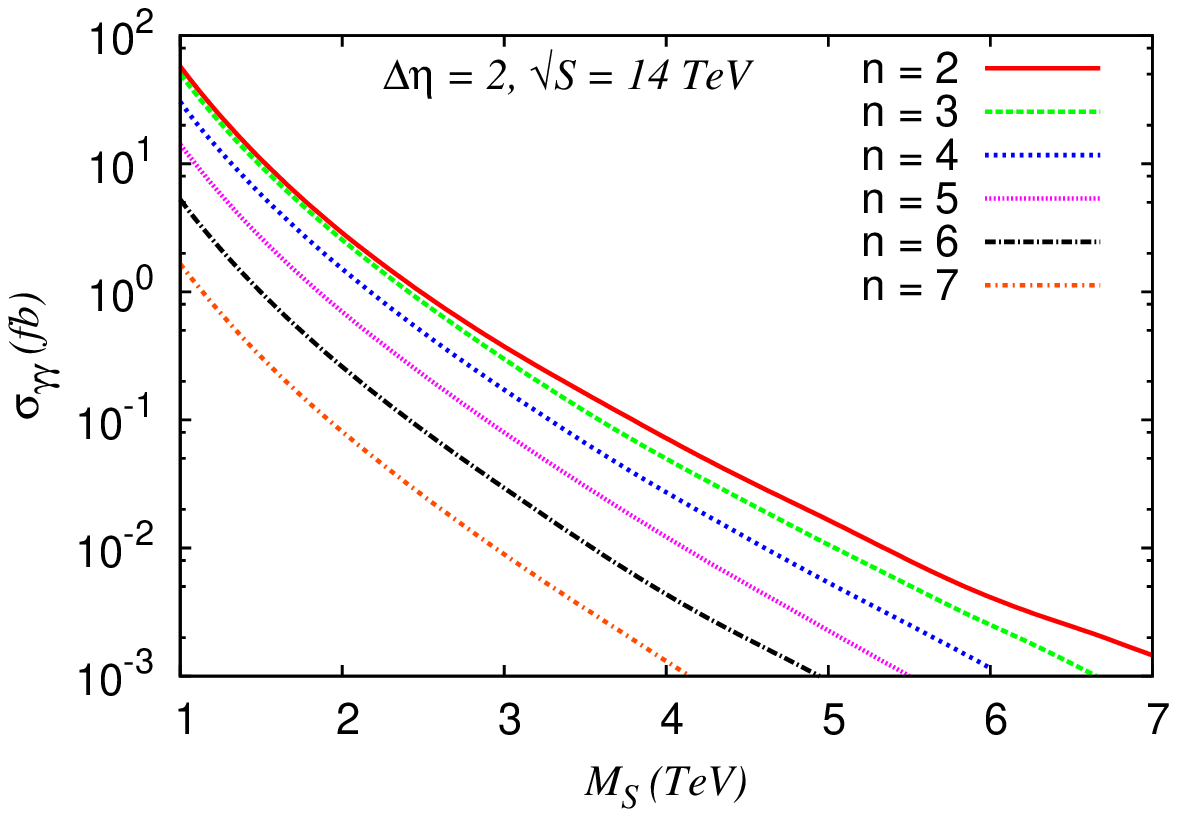}} \\
\footnotesize{\hspace{1.1cm}(a)}&&\footnotesize{\hspace{1.2cm}(b)}
\end{tabular}}
\caption{\label{fig:tCS_Ms_YY} The total signal cross sections ($pp\to X+\gamma\gamma+Y$) in diphoton
channel ($\sigma_{\gamma\gamma}$) as a function of $M_S$ at the 14 TeV LHC. In (a) we plot $\sigma_{\gamma\gamma}$ for $\Delta\eta=$ 1.5, 2.0, 2.5 and 3.0 for $n=2$. In (b) we plot $\sigma_{\gamma\gamma}$ for $n=2-7$ for $\Delta\eta=2$.
In both plots $\sigma_{\gamma\gamma}$'s are computed after applying the cut defined in Eq. \ref{eq:tCS_cuts1}.}
\end{figure}

In Fig. \ref{fig:tCS_sqrtS_YY} we show the total signal cross section in the 
diphoton channel ($\sigma_{\gamma\gamma}$) as a function of $\sqrt{S}$ after applying the cut defined in Eq. \ref{eq:tCS_cuts1}.
In Fig. \ref{fig:tCS_sqrtS_YY}(a) we show $\sigma_{\gamma\gamma}$ for different $\Delta\eta$
taking $n=2$ and $M_S=2$ TeV. In Fig. \ref{fig:tCS_sqrtS_YY}(b) we show $\sigma_{\gamma\gamma}$ for different $n$ varied from two to seven.
In Fig. \ref{fig:tCS_Ms_YY} we show the $\sigma_{\gamma\gamma}$ as a function of $M_S$ at the 14 TeV LHC.
In Fig. \ref{fig:tCS_Ms_YY}(a) we show $\sigma_{\gamma\gamma}$ for different $\Delta\eta$
taking $n=2$ and in Fig. \ref{fig:tCS_Ms_YY}(b) we show $\sigma_{\gamma\gamma}$ for different $n$ taking $\Delta\eta=2$. Here we compute the $\sigma_{\gamma\gamma}$
after applying the kinematical cuts defined in Eq. \ref{eq:tCS_cuts1}. 
The behavior of these plots are very similar to the corresponding plots for
dilepton channel and we are not repeating those features here. An interesting point
to notice that $\sigma_{ll}$ and $\sigma_{\gamma\gamma}$ are similar even 
quantitatively. This is because surprisingly $\sigma_{ll}$ and $\sigma_{\gamma\gamma}$ are equal after $\cos\theta$ integration from $-1$ to $1$
as follows:
\begin{equation}
\sum_{l=e,\mu}\int^{1}_{-1}d(\cos\theta)\frac{d\hat\sigma(gg\rightarrow l^{+}l^{-})}{d|\cos\theta|} = \int^{1}_{-1}d(\cos\theta)\frac{d\hat\sigma(gg\rightarrow \gamma\gamma)}{d|\cos\theta|}
\end{equation}

\begin{figure}[!h]
\centering
\subfloat{
\begin{tabular}{ccc}
\resizebox{80mm}{!}{\includegraphics{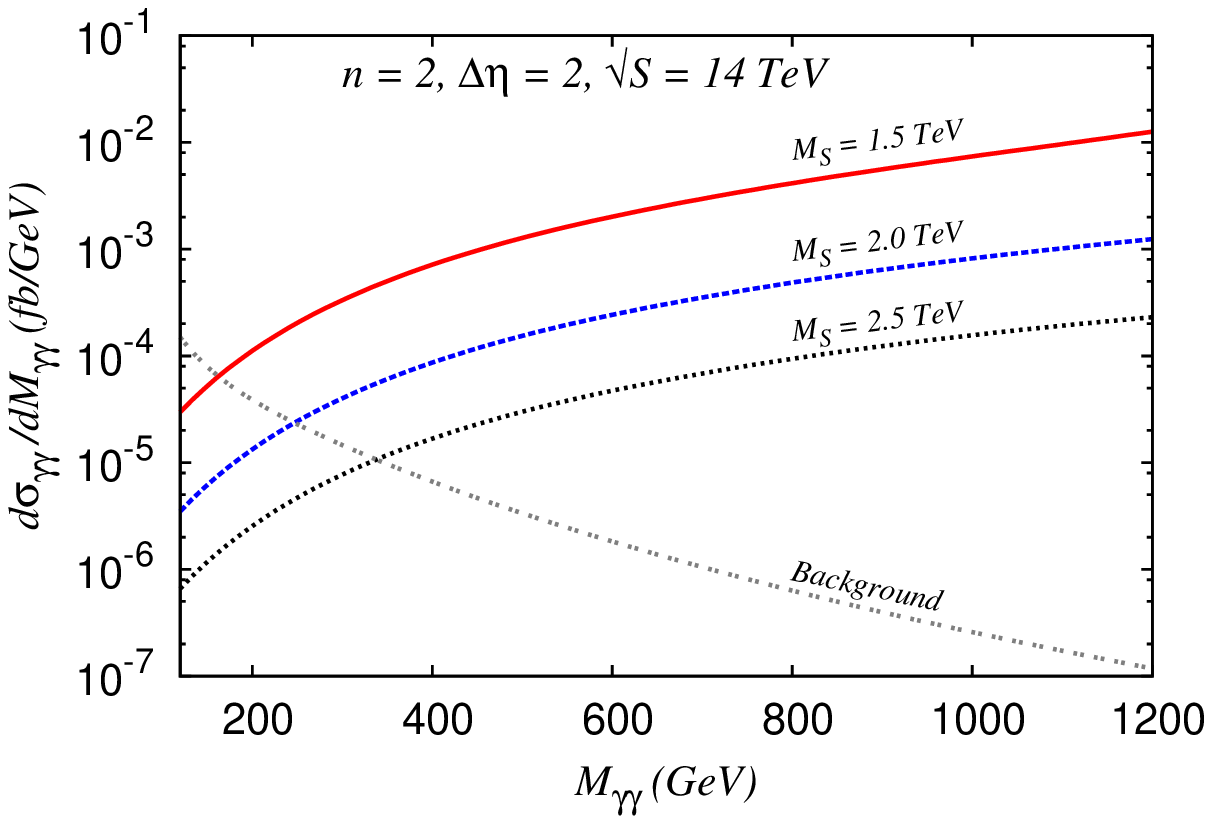}} &&
\resizebox{80mm}{!}{\includegraphics{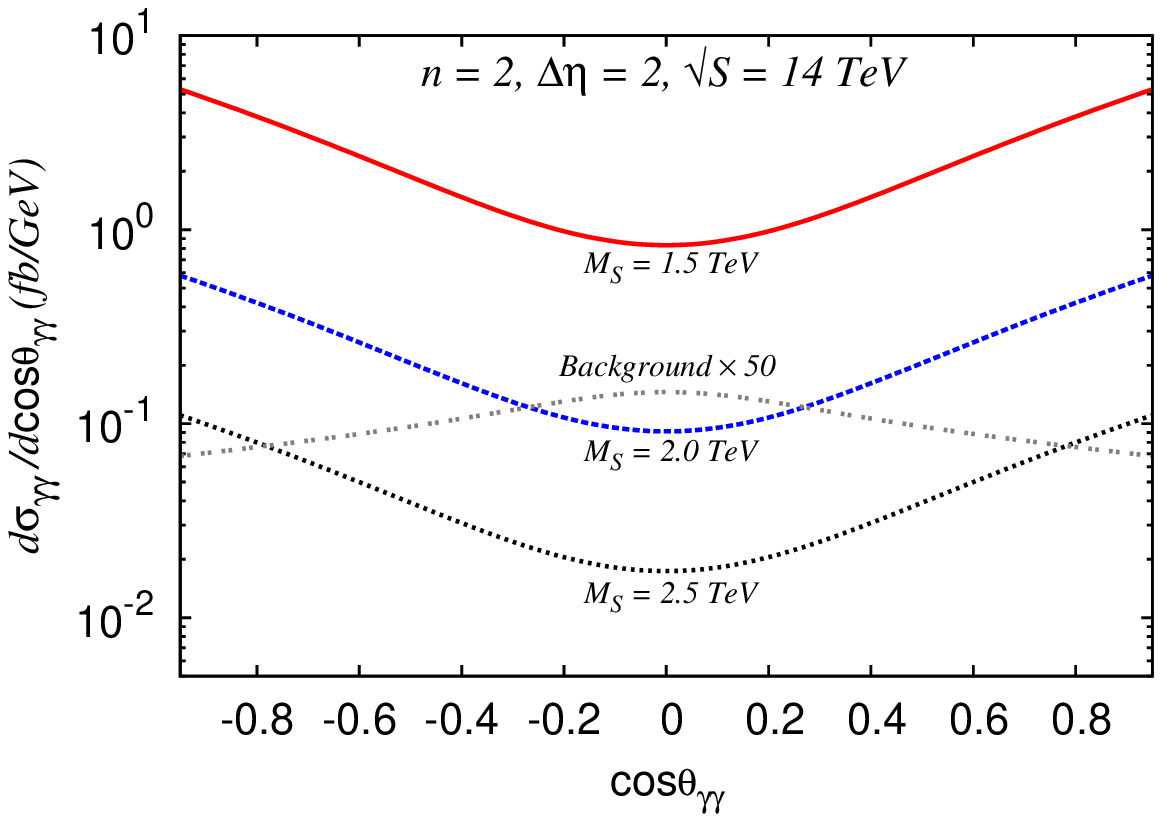}}\\
\footnotesize{\hspace{1.1cm}(a)}&&\footnotesize{\hspace{1.2cm}(b)}
\end{tabular}}
\caption{\label{fig:dCS_YY} (a) The invariant mass distributions ($d\sigma_{\gamma\gamma}/dM_{\gamma\gamma}$) for signal and background ($pp\to X+\gamma\gamma+Y$) of the photon pair after applying the cuts defined in Eq. \ref{eq:dCS_cuts1} at the 14 TeV LHC. (b) The angular distributions ($d\sigma_{\gamma\gamma}/d\cos\theta_{\gamma\gamma}$) for signal and background after applying the cuts defined in Eq. \ref{eq:dCS_cuts2} at the 14 TeV LHC. For both plots we take $M_S=$ 1.5, 2.0 and 2.5 TeV and $n=2$ for signal and, $\Delta\eta=2$ for
both signal and 
background. The background in (b) is shown after multiplying by a factor of 50.}
\end{figure}

In Fig. \ref{fig:dCS_YY}(a) we show signal and background invariant mass 
distributions ($d\sigma_{\gamma\gamma}/dM_{\gamma\gamma}$) of the diphoton pair 
after applying the selection cuts defined in Eq. \ref{eq:dCS_cutsYY1} at the 14 TeV LHC.
\begin{equation}
\label{eq:dCS_cutsYY1}
|y_{\gamma\gamma}| \leq 2;~~|\cos\theta_{\gamma\gamma}|\leq 1
\end{equation}
We plot signal 
$d\sigma_{\gamma\gamma}/dM_{\gamma\gamma}$ for $M_S=$ 1.5, 2.0 and 2.5 TeV taking $n=2$ and,
$\Delta\eta=2$ both for signal and background. The signal $d\sigma_{\gamma\gamma}/dM_{\gamma\gamma}$ is a
monotonically increasing function of $M_{\gamma\gamma}$ ($M_{\gamma\gamma}^2=\hat{s}$) because the subprocess cross section
increases roughly as $\hat{s}^3$ (see Eq. \ref{eq:dCSdcos_YY}). The background cross section is computed including $gg$ and $\gamma\gamma$ initiated
processes as shown in Fig. \ref{fig:dipho_bkgrnd}. Although the background in Fig.
\ref{fig:dCS_YY}(a) can beat signal in the small $M_{\gamma\gamma}$ region, it 
dies out very rapidly in the high $M_{\gamma\gamma}$ region. This is because the
diphoton background goes roughly as $1/M_{\gamma\gamma}^2$ and becomes almost
negligible at large $M_{\gamma\gamma}$. In Fig. \ref{fig:dCS_YY}(b) we show signal and background angular
distributions ($d\sigma_{\gamma\gamma}/d\cos\theta_{\gamma\gamma}$) of the diphoton pair after applying the selection cuts defined in Eq. \ref{eq:dCS_cutsYY2} at the 14 TeV LHC.
\begin{equation}
\label{eq:dCS_cutsYY2}
|y_{\gamma\gamma}| \leq 2;~~250~\textrm{GeV}  \leq M_{\gamma\gamma} < 1250~\textrm{GeV} 
\end{equation}

We see that diphoton signal and background angular distributions behave just in the opposite way to the dilepton signal and background angular distributions respectively. The signal $d\sigma_{\gamma\gamma}/d\cos\theta_{\gamma\gamma}
\sim (1+6\cos^2\theta+\cos^4\theta)$, has
a minimum at $\cos\theta=0$ and attains its maximum value at $|\cos\theta|=1$.
Whereas, the $d\sigma_{\gamma\gamma}/d\cos\theta_{\gamma\gamma}$ for the 
dominant $gg\to \gamma\gamma$ background attains its maximum value at $\cos\theta=0$ 
as shown in Fig. \ref{fig:dCS_YY}(b).
This is because of the nature of angular dependent part in Eq. \ref{eq:dCSdcos_bkgrnd}. 
The similar angular behavior is true for subdominant $\gamma\gamma\to\gamma\gamma$
background also. Whereas, there is no angular dependency of other subdominant 
$gg\to h \to \gamma\gamma$ backgrounds because of the spin-0 nature of the intermediate Higgs.

\subsection{LHC Discovery Potential}
We define the luminosity requirement for the discovery of KK gravitons 
at the LHC as following:
\begin{equation}
L_{D} =  Max\{L_5,L_{10}\}
\end{equation}
where $L_5$ denotes the luminosity required to attain $5\sigma$ statistical
significance and $L_{10}$ is the luminosity required to
observe 10 signal events. We compute $L_D$ after applying some kinematical cuts
which we call the ``Discovery Cuts'' as defined below.

\begin{figure}[!h]
\centering
\subfloat{
\begin{tabular}{ccc}
\resizebox{80mm}{!}{\includegraphics{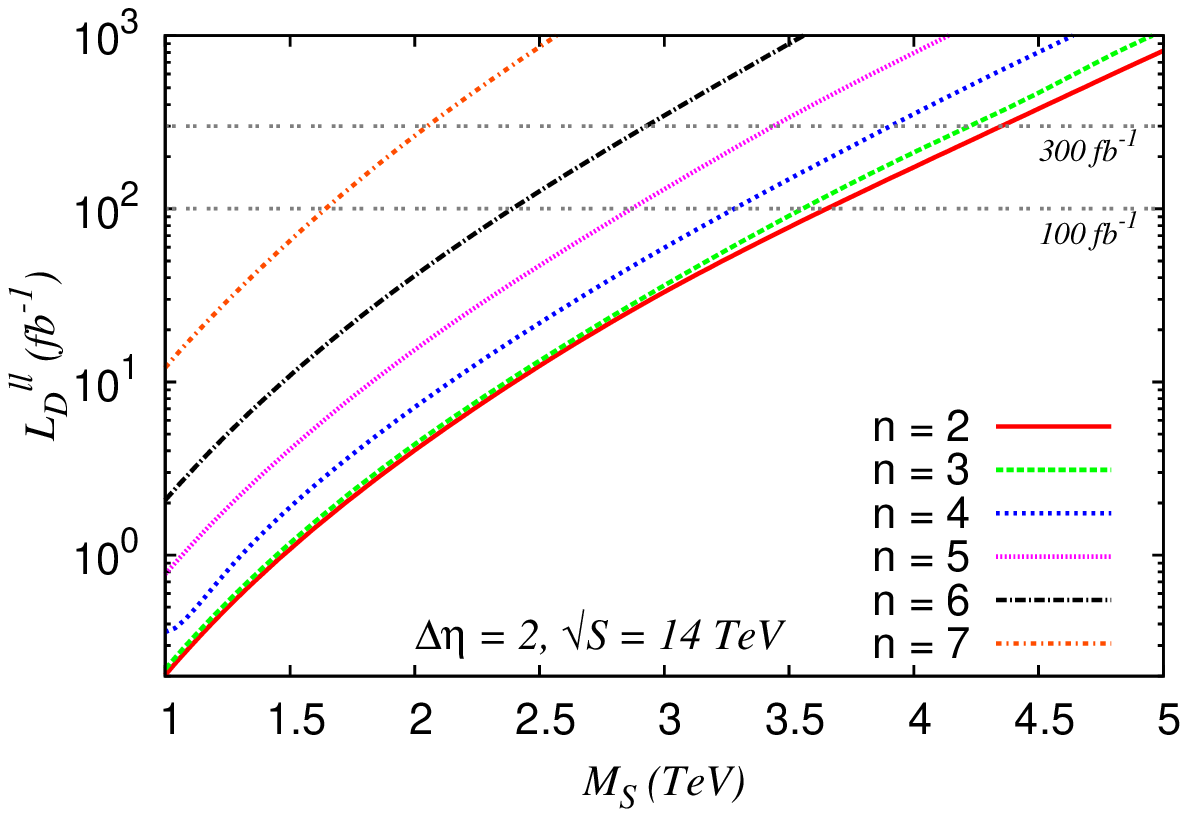}} &&
\resizebox{80mm}{!}{\includegraphics{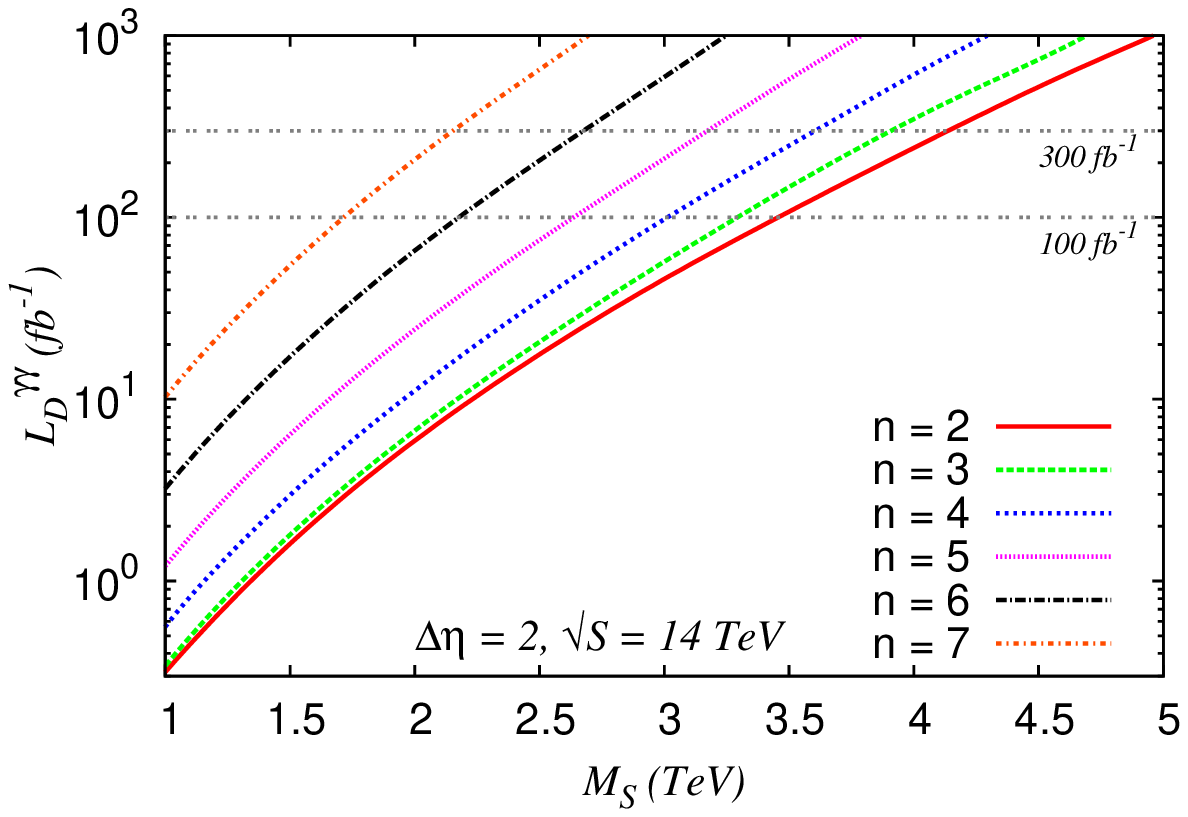}}\\
\footnotesize{\hspace{1.1cm}(a)}&&\footnotesize{\hspace{1.2cm}(b)}
\end{tabular}}
\caption{\label{fig:LD_LL_YY} The required luminosity ($L_D$) for the discovery of KK graviton in (a)
dilepton and (b) diphoton channels as a function of $M_S$ at the 14 TeV LHC with ``Discovery cuts'' (see text for the definitions of the cuts). The $L_D^{ll}$ and
$L_D^{\gamma\gamma}$ are computed for $n=2-7$ with $\Delta\eta=2$ at
$pp\to X+\psi\psi+Y$ level where $\psi=\{l,\gamma\}$ .}
\end{figure}    

\subsection*{Discovery cuts}
Motivated by the invariant mass and angular distributions in Figs. \ref{fig:dCS_LL}
and 
\ref{fig:dCS_YY} we construct some kinematical cuts to separate the signal from the background in the dilepton and diphoton channels as follows: 
\begin{enumerate}
\item Dilepton channel: $|y_{ll}| \leq 2;~~|\cos\theta_{ll}|\leq 0.8;~~M_S/2 \leq M_{ll} < M_S - 10~\textrm{GeV}$

\item Diphoton channel: $|y_{\gamma\gamma}| \leq 2;~~|\cos\theta_{\gamma\gamma}|\leq 1;~~250~\textrm{GeV} \leq M_{\gamma\gamma} < M_S - 10~\textrm{GeV}$
\end{enumerate}
We call these cuts as the ``Discovery cuts'' for the discovery of KK graviton
at the LHC.
In Fig. \ref{fig:LD_LL_YY}, $L_D$ goes as $L_{10}$ for both dilepton and diphoton
channels in the whole parameter space we have displayed. In case of the diphoton channel the background is already quite small compared to the signal. Whereas, for the dilepton channel, background becomes under
control after the ``Discovery Cuts''. With the ``Discovery cuts'' one can probe 
$M_S$ roughly up to 3.6 TeV (4.4 TeV) with $100$ fb$^{-1}$ ($300$ fb$^{-1}$) 
integrated luminosity at the 14 TeV LHC.

\section{Conclusions}
\label{sec:conclusion}

In this paper we have discussed the possibility of looking for new 
physics, in particular the signatures of large extra dimensions, in diffractive
processes at the LHC. We have concentrated on the central inclusive
production where the outgoing protons breakup, but the fragments are
mostly emitted in the forward direction. We have studied in detail the signal
of KK gravitons via the dilepton and the diphoton channels. From angular
momentum conservation, the spin-2 graviton cannot be produced at LO in exclusive
processes as discussed in the text, but it can be produced in an
inclusive configuration. Although the backgrounds are larger for inclusive
production, the larger cross section and event rate for signals make the collider
search for 
extra dimensional theories very promising. Since the intermediate particle
is spin-2 in nature, we also get a unique angular distribution which can be
used to distinguish signals from backgrounds. We have shown that it is very unlikely
to produce a lepton or a photon pair in the SM with very high invariant mass in the
inclusive configuration.
Thus, experimentally if we find  a reasonable number of lepton or photon pairs with high invariant mass, it would
clearly indicate the signature of new physics possibly the signature of large
extra dimensions.

In this work, we have indicated the possibility of using dilepton and
diphoton in the final state to signal the presence of extra dimensions. We have carefully estimated the sizes of the signal and the
background and, found that the signal rates are not too big. Nevertheless, it is possible within
reasonable parameter ranges as suggested by present day studies, to use 
central production, with forward detectors for the proton fragments, to look 
for KK gravitons as a signal for ADD like theories.
The great advantage of using such channels for the study
of new physics is that they are clean and the signals are easily
distinguished from the backgrounds. On the flip side, the event rates are
considerably smaller than other standard new physics searches. This can be compensated by runs at higher energies and greater
luminosities at some future date. There is another general issue with high
luminosity run at the LHC - the presence of huge ``pile-up'' background. This
background can overwhelm the rapidity gap signatures. If we consider
inclusive processes and sacrifice the forward proton tagging (applicable only for
exclusive processes), we can detect
rapidity gap events at the LHC \cite{Ryutin:2012np}.  

All higher 
dimensional models are valid up to a cutoff scale. In our case, the signal cross sections depend 
on the eighth power of $1/M_S$ ($M_S$ is the cutoff scale of the theory). Thus, numerical results are very sensitive 
to the UV cutoff. The experimental lower bound of $M_S\approx 3$ TeV for $n=2$. 
Future experiments can push up the lower bound and make the signatures less 
visible. We show that LHC with $\sqrt{S}=14$ TeV and 100 fb$^{-1}$
(300 fb$^{-1}$) of integrated luminosity can probe $M_S$
up to 3.6 TeV (4.4 TeV) in both the dilepton and the diphoton channels.

\section*{Acknowledgments}
We are grateful to Subhadip Mitra for many fruitful discussions and assistance in certain numerical aspects of this work.

\appendix

\section{Inclusive Luminosity}
\label{app:inclu_lum}

\begin{figure}[!h]
\centering
\includegraphics[scale=0.9]{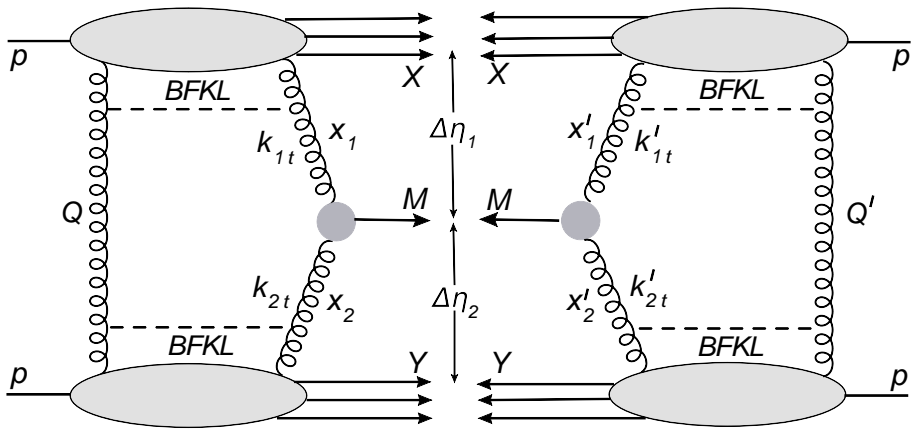}
\caption{\label{fig:inclu_MM} The inclusive double diffractive amplitude multiplied by its complex conjugate. A central system $M$ is produced with rapidity gaps
$\Delta\eta_1$ and $\Delta\eta_2$ between proton dissociated products $X$ and $Y$ respectively.}
\end{figure}

An outline of the derivation of effective luminosity for the inclusive 
configuration is presented in \cite{Khoze:2001xm}. Here, we briefly give some
useful formulae for computing inclusive luminosity.
At the partonic level the inclusive process in Fig. \ref{fig:inclu_MM} can be expressed as the exclusive
production of the system $M$ with rapidity gaps on either side i.e.,
\begin{equation}
a_1a_2 \to a_1 + M + a_2.
\end{equation}

The probabilities to find partons $a_i$ ($i=\{1,2\}$) in the protons is given by the 
effective parton densities $\mathcal{G}(x_i,k^2_{it})$ evaluated
at scale $k^2_{it}$ with momentum fraction $x_i$ as given by
\begin{equation}
\mathcal{G}(x_i,k^2_{it}) = x_ig(x_i,k^2_{it}) + \frac{16}{81}\sum_{q}x_i\left[q(x_i,k^2_{it}) +\bar{q}(x_i,k^2_{it})\right].
\end{equation}
where $g$, $q$ and $\bar{q}$ in the above equation are the usual PDF's.
The rapidity gaps can be filled up due to gluon bremsstrahlung effect. The mean number of gluons emitted
in transverse momentum interval, $Q_t < P_t < k_{it}$ (also $Q_t^2  \ll k_{it}^2$, asymmetric $t$-channel gluon exchange) with rapidity gap $\Delta\eta_i$ is given by
\begin{equation}
\label{eq:mean_glu}
n_i = \frac{3\alpha_S}{\pi}\Delta\eta_i\ln\left(\frac{k_{it}^2}{Q^2_t}\right).
\end{equation}
Therefore, the amplitude for no emission in the rapidity interval $\Delta\eta_i$
can be expressed as $A_i = \exp\left(-n_i/2\right)\Phi(Y_i)$. The amplitude $A_i$ 
is called the non-forward BFKL amplitude which is computed resumming the 
double logarithms in Eq. \ref{eq:mean_glu}. The factor $\Phi(Y_i)$ accounts for
the longitudinal BFKL logarithm where $Y_i=(3\alpha_S/2\pi)\Delta\eta_i$. For
rapidity gap $\Delta\eta_i \lesssim 4$ we have $Y_i\lesssim 0.4$. In the asymmetric 
region, $Q_t^2  \ll k_{it}^2$, it is sufficient to keep only the $\mathcal{O}(Y_i)$
term, i.e. $\Phi(Y_i)\approx 1 + Y_iQ_t^2/k_{it}^2 \approx 1.1 \pm 0.1$
\cite{Forshaw:1995ax}. There is another suppression comes in the form of 
Sudakov factor $T(k_{it},\mu)$ which accounts for the survival probability 
of a gluon in the interval $k_{it} < P_t < \mu$ where $\mu$ is related to the hard
scale as $\mu = M/2$. The Sudakov factor reads as 
\begin{equation}
T\left(k_{it},\mu \right) = \exp\left(-\int^{\mu^2}_{k_{it}^2}
\frac{\alpha_S(k_t^2)}{2\pi}\frac{dk^2_t}{k_t^2}
\int^{1-\Delta}_0\left[zP_{gg}(z)+\sum_q P_{qg}(z)\right]dz\right)
\end{equation}
where $P_{gg}(z)$ and $P_{qg}(z)$ are the LO DGLAP splitting functions
and $\Delta =  k_t/(\mu + k_t)$. We obtain the effective luminosity ($\mathcal{L}_{gg}$) by multiplying the inclusive amplitude by its complex conjugate as shown in Fig. \ref{fig:inclu_MM}. To compute $\mathcal{L}_{gg}$ we carefully
account for all the factors i.e. non-forward BFKL amplitudes, Sudakov factors, gap
survival factor and appropriate color factors, and finally arrive at the expression 
\begin{eqnarray}
\mathcal{L}_{gg} = \mathcal{S}^2\int^{1}_{x_1^{min}}\mathcal{G}(x_1,k_{1t}^2)\frac{dx_1}{x_1}\int^{1}_{x_2^{min}}\mathcal{G}(x_2,k_{2t}^2)\frac{dx_2}{x_2}\frac{\alpha_S^4}{\pi^2}
\left(\frac{N_c^2}{N_c^2-1}\right)^2\mathcal{I}_{gg} 
\end{eqnarray}
\begin{eqnarray}
\textrm{where}~~\mathcal{I}_{gg} = \int\frac{dQ_t^2}{Q_t^2}\frac{d{Q^{\prime}}_t^2}{{Q^{\prime}}_t^2}
\frac{dk_{1t}^2}{k_{1t}^2}\frac{dk_{2t}^2}{k_{2t}^2}
\left(A_1A_2A^{\prime}_1A^{\prime}_2\right)\sqrt{T_1T_2T^{\prime}_1T^{\prime}_2}
\end{eqnarray}
Here all the primed quantities are arising from the complex conjugate of the 
amplitude.
The minimum of the momentum fraction $x_i^{min}$ maintains the rapidity 
gap $\Delta\eta_i$ and  given by
\begin{equation}
x_i^{min} = \frac{M}{\sqrt{S}}\exp({y}) + \frac{k_{it}}{\sqrt{S}}\exp({y + \Delta\eta_i})
\end{equation}

The momentum of the screening gluon $Q$ is very small. Therefore, in the limit
$Q^2\ll k_i^2$, we have $t_i = (Q-k_i)^2\approx -k_{it}^2\approx 
-{k^{\prime}_{it}}^2$.
After performing $Q_t^2$ and ${Q^{\prime}_t}^2$ integrations $\mathcal{I}_{gg}$
takes the simplified form as follows
\begin{equation} 
\mathcal{I}_{gg} = \frac{1}{(Y_1 + Y_2)^2}\int\frac{dt_1}{t_1}\frac{dt_2}{t_2}
\exp\left( - \frac{3\alpha_S}{\pi}\Delta\eta\left|\ln\frac{t_1}{t_2}\right|\right)
T\left(\sqrt{|t_1|},\mu\right)T\left(\sqrt{|t_2|},\mu\right)
\end{equation}
where $\Delta\eta=\Delta\eta_1$ if $|t_1|>|t_2|$, but $\Delta\eta=\Delta\eta_2$ 
if $|t_1|<|t_2|$. In our computation of $\mathcal{L}_{gg}$ we have used this simplified form of $\mathcal{I}_{gg}$ taking $\Delta\eta_1=\Delta\eta_2\equiv \Delta\eta$.

\section{Feynman Rules and Graviton Propagator}
\label{app:Feynrules}

In this appendix we give the Feynman rules for $VVG$ (where $V$ is a vector boson with mass $M_V$) and $ffG$ (where $f$ is a fermion with mass $M_f$) vertices
\cite{Han:1998sg}
\begin{eqnarray}
\begin{tikzpicture}
\node[left] at (0,0) {\includegraphics[width=3.5cm]{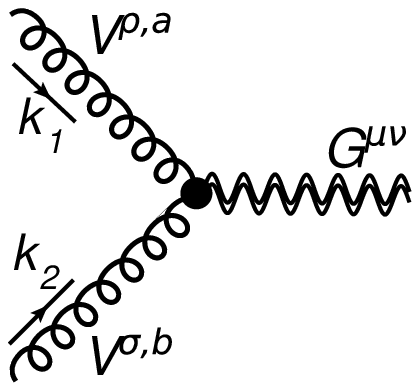}};
\node[] at (0,0) { $\quad\equiv$ };
\node[right] at (0,0) { $\quad \displaystyle -i\frac{\kappa}{2}\delta^{ab}\left[(M_V^2+k_1\cdot k_2)C_{\mu\nu,\rho\sigma} + D_{\mu\nu,\rho\sigma}(k_1\cdot k_2) + \frac{1}{\xi}E_{\mu\nu,\rho\sigma}(k_1\cdot k_2)\right]$};
\end{tikzpicture}
\end{eqnarray}
\begin{eqnarray}
\begin{tikzpicture}
\node[left] at (0,0) {\includegraphics[width=3.5cm]{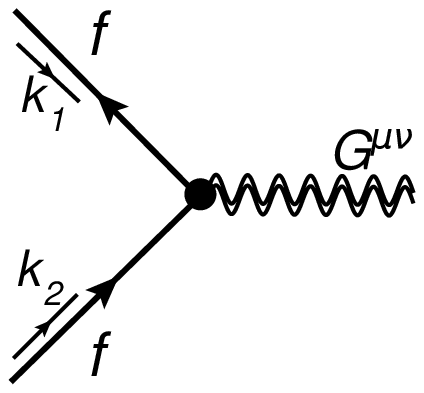}};
\node[] at (0,0) { $\quad\equiv$ };
\node[right] at (0,0) { $\quad \displaystyle -i\frac{\kappa}{8}\left[\gamma_{\mu}(k_1 - 
k_2)_{\nu}+\gamma_{\nu}(k_1 - k_2)_{\mu} - 2\eta_{\mu\nu}
(\kern+0.30em 
/\kern-0.70em{k_1}-\kern+0.30em
/\kern-0.70em{k_2}-2M_f)\right]$};
\end{tikzpicture}
\end{eqnarray}
where $\kappa=\sqrt{16\pi G_N}$. The $G_N=(4\pi)^{n/2}/R^{n}M_S^{2+n}$ is the $(4+n)$ dimensional Newton constant.
 All the tensors appear in the Feynman rules are defined as
\begin{equation}
C_{\mu\nu,\rho\sigma} = \eta_{\mu\rho}\eta_{\nu\sigma} + \eta_{\mu\sigma}\eta_{\nu\rho} - \eta_{\mu\nu}\eta_{\rho\sigma}~~\textrm{where}~~
\eta_{\mu\nu} = \textrm{diag}\{+1,-1,-1,-1\}
\end{equation}
\begin{eqnarray}
D_{\mu\nu,\rho\sigma}(k_1\cdot k_2) = \eta_{\mu\nu}k_{1\sigma}k_{2\rho}
-\left[\eta_{\mu\sigma}k_{1\nu}k_{2\rho} +\eta_{\mu\rho}k_{1\sigma}k_{2\nu} 
-\eta_{\rho\sigma}k_{1\mu}k_{2\nu} + \left( \mu \leftrightarrow \nu\right)\right]
\end{eqnarray}
\begin{eqnarray}
E_{\mu\nu,\rho\sigma}(k_1\cdot k_2) = \eta_{\mu\nu}\left(
k_{1\rho}k_{1\sigma} +k_{2\rho}k_{2\sigma} +k_{1\rho}k_{2\sigma} \right) - 
\left[\eta_{\nu\sigma}k_{1\mu}k_{1\rho} +\eta_{\nu\rho}k_{2\mu}k_{2\sigma} +\left( \mu \leftrightarrow \nu\right)\right]
\end{eqnarray}

The propagator for spin-2 KK graviton with mass $M_G$ and decay width 
$\Gamma_G$ is given by
\begin{eqnarray}
\Delta_{\mu\nu,\rho\sigma}(k) = \frac{\frac{i}{2}B_{\mu\nu,\rho\sigma}(k)}{k^2 - M_G^2 + i\Gamma_G M_G}~~\textrm{where}~~
\end{eqnarray}
\begin{eqnarray}
B_{\mu\nu,\rho\sigma}(k) &=&
\left(\eta_{\mu\rho} - \frac{k_{\mu}k_{\rho}}{M_G^2}\right)
\left(\eta_{\nu\sigma} - \frac{k_{\nu}k_{\sigma}}{M_G^2}\right) +
\left(\eta_{\mu\sigma} - \frac{k_{\mu}k_{\sigma}}{M_G^2}\right)
\left(\eta_{\nu\rho} - \frac{k_{\nu}k_{\rho}}{M_G^2}\right)\nonumber \\
&-&\frac{2}{3}
\left(\eta_{\mu\nu} - \frac{k_{\mu}k_{\nu}}{M_G^2}\right)
\left(\eta_{\rho\sigma} - \frac{k_{\rho}k_{\sigma}}{M_G^2}\right)
\end{eqnarray}

Mass of the $k$-th KK mode is $M_k = 4\pi^2k^2/R^2$ and mass separation between 
two adjacent KK modes is a $\mathcal{O}(1/R)$ term. Therefore, KK modes become
quasi-continuous and it is convenient to define KK state density. The number
of KK states in the mass interval $M_k^2$ and $M_k^2+dM_k^2$ for $4+n$ dimensional space can be expressed as
$\Delta k^2= \rho(M_k)dM_k^2$ where
\begin{equation}
\rho(M_k) = \frac{R^nM_k^{n-2}}{(4\pi)^{n/2}\Gamma(n/2)}.
\end{equation}

The effective interaction due to all KK states contributing to a physical 
process can be obtained after summing over all the propagators as follows
\begin{equation}
\mathcal{D}(\hat{s}) = \sum_k \frac{i}{\hat{s} - M_k^2 + i\varepsilon}=
\int^{\infty}_0dM_k^2~\rho(M_k)\frac{i}{\hat{s} - M_k^2 + i\varepsilon}
\end{equation}
where $\varepsilon=M_G\Gamma_G$ and $\hat{s}=k^2$. The effective propagator $\mathcal{D}(\hat{s})$ valid up to $M_S$ looks 
\begin{equation}
\label{eq:exact_propagator}
\mathcal{D}(\hat{s}) = \frac{\hat{s}^{n/2-1}R^{n}}{\Gamma(n/2)(4\pi)^{n/2}}\left[\pi + 2iI\left(\frac{M_S}{\sqrt{\hat{s}}}\right)\right]~~\textrm{where}~~
\end{equation}
\begin{equation}
I\left(\frac{M_S}{\sqrt{\hat{s}}}\right) = \left\{ 
\begin{array}{l l}
- \displaystyle\sum_{k=1}^{n/2-1}\frac{1}{2k}
\left(\frac{M_S}{\sqrt{\hat{s}}}\right)^{2k} - \frac{1}{2}\ln\left(\frac{M_S^2}{\hat{s}} -1\right) & \quad \mbox{for}~ n=\mbox{even}\\
- \displaystyle\sum_{k=1}^{(n-1)/2}\frac{1}{2k-1}
\left(\frac{M_S}{\sqrt{\hat{s}}}\right)^{2k-1} + \frac{1}{2}\ln\left(\frac{M_S+\sqrt{\hat{s}}}{M_S-\sqrt{\hat{s}}}\right) & \quad \mbox{for}~ n=\mbox{odd}\\ \end{array} \right.
\end{equation}

In the limit $M_S^2\gg s$, the effective propagator $\mathcal{D}(\hat{s})$ becomes
\begin{equation}
\label{eq:apprx_Ds}
|\mathcal{D}(\hat{s})| = \frac{16\pi}{\kappa^2 M_S^4}\mathcal{K}~~\textrm{where}~~
\mathcal{K}= \left\{
\begin{array}{l l}
\ln\left(M_S^2/\hat{s}\right) & \quad \mbox{for}~ n=\mbox{2}\\
2/(n-2) & \quad \mbox{for}~ n>\mbox{2}\\ \end{array} \right.
\end{equation}

In our analysis we have used the exact form of $\mathcal{D}(\hat{s})$ given in Eq. \ref{eq:exact_propagator}.
   
\section{Dilepton Background}
\label{app:dilep_bkgrnd}

\begin{figure}[!h]
\centering
\includegraphics[scale=0.8]{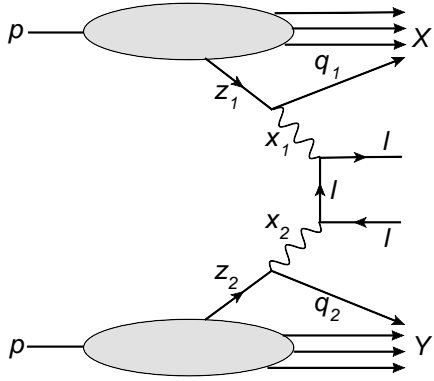}
\caption{\label{fig:inelastic_YY} Inelastic photo-production of a lepton pair at the LHC. The photons are coming from the quarks inside protons.}
\end{figure} 

We consider the inelastic photo-production of a lepton pair at the LHC shown in Fig. \ref{fig:inelastic_YY} as follows
\begin{equation}
pp\to X+\gamma\gamma + Y \to X+l^+l^- +Y
\end{equation}

A photon with momentum fraction $z_1$ ($z_2$) w.r.t. the quark momentum is emitted from the quark $q_1$ ($q_2$) with momentum fraction $x_1$ ($x_2$) of a proton.
The total inelastic $pp\to l^+l^-$ cross section can be computed (this has been thoroughly computed in \cite{Drees:1994zx}) using Weizs\"{a}cker-Williams equivalent photon approximation \cite{von Weizsacker:1934sx,Williams:1934ad} as given by
\begin{eqnarray}
\sigma(S) &=& \sum_{q_1,q_2}\int^{1}_{\frac{4m_l^2}{S}}dx_1
\int^{1}_{\frac{4m_l^2}{Sx_1}}dx_2\int^{1}_{\frac{4m_l^2}{Sx_1x_2}}dz_1
\int^{1}_{\frac{4m_l^2}{Sx_1x_2z_1}}dz_2~e^2_{q_1}e^2_{q_2} \nonumber \\
&\times & f_{q_1/p}\left(x_1,Q^2\right)f_{q_2/p}\left(x_2,Q^2\right)
f_{\gamma/q_1}(z_1)f_{\gamma/q_2}(z_2)\hat{\sigma}(\hat{s})
\end{eqnarray}
where $q_1,q_2$ are the quarks (we denote $q\equiv\{q_1,q_2\}$), $m_l$ is the mass of the lepton, $e_q$ is the EM charge of $q$ in the unit of an electron's charge and $\hat{\sigma}$ is the subprocess cross section with CM energy $\sqrt{\hat{s}}$ ($\hat{s} = x_1x_2z_1z_2S$, $S$ is the LHC CM energy). The quark 
density inside the proton is denoted as $f_{q/p}$. The $f_{\gamma/q}$ is the photon spectrum inside a quark and is given by
\begin{equation}
f_{\gamma/q}(z) =
\frac{\alpha_{em}}{2\pi}\frac{\left[1+(1-z)^2\right]}{z}\ln\left(\frac{Q_1^2}{Q_2^2}\right)
\end{equation}
We choose the scale $Q_1=\sqrt{\hat{s}}/2$ and $Q_2=1$ GeV. The subprocess cross section for $\gamma\gamma\to ll$ is given by
\begin{equation}
\hat{\sigma} = \frac{4\pi\alpha^2_{em}(M_W^2)}{\hat{s}}
\left[\frac{3-\beta^4}{2}\ln\left(\frac{1+\beta}{1-\beta}\right)
-2\beta +\beta^3\right]
\end{equation}
where $\beta = \sqrt{1-4m_l^2/\hat{s}}$. We have used $\alpha_{em}=1/137$ and
$\alpha^2_{em}(M_W^2)=1/128$ in our analysis.

The differential cross section of the subprocess in the high energy limit 
is 
\begin{equation}
\label{eq:dCSdcosL_bkgrndLL}
\frac{d\hat{\sigma}}{d|\cos\theta|} = \frac{2\pi\alpha^2_{em}(M_W^2)}{\hat{s}}
\left(\frac{1+\cos^2\theta}{\sin^2\theta}\right)
\end{equation}
where $\theta$ is scattering angle of the lepton in the CM frame. We see $d\hat{\sigma}/d|\cos\theta|\to \infty$ if $|\cos\theta|\to 1$.

\section{Diphoton Background}
\label{app:bkgrnd_YY}
We compute the dominant background cross section for the diphoton channel $gg\to \gamma\gamma$  via a quark loop  using the 
$\gamma\gamma\to \gamma\gamma$ helicity amplitudes given in \cite{Jikia:1993tc}.
The differential cross section of $gg\to \gamma\gamma$ background in the high energy limit and for small $\theta$ (scattering angle of
$\gamma$ in the CM frame) is given by
\begin{equation}
\label{eq:dCSdcos_bkgrnd}
\frac{d\sigma}{d|\cos\theta|} =\frac{1}{2}\cdot\left(\frac{1}
{32\pi\hat{s}}\right)\cdot\left(\frac{1}{8^2}\right)\cdot\left(\frac{1}{2^2}\right)\cdot 8\cdot
N_fN_c\cdot\alpha_S^2\alpha_{em}^2|\mathcal{M}|^2
\end{equation}
Origin of some factors in the above equation are once explained in sec. \ref{sebsec:diphoton} and we are not explaining them
again.
The factor $N_fN_c$ is the number of possible quarks that can contribute in the loop. 
The $|\mathcal{M}|^2$ is the sum of two helicity amplitudes, which are the
only two surviving amplitudes in the high energy limit,
\begin{equation}
|\mathcal{M}|^2 = |\mathcal{M}_{++++}|^2 + |\mathcal{M}_{+-+-}|^2 = 32\ln^4\left(\frac{2}{1+|\cos\theta|}\right).
\end{equation}

\end{document}